\documentclass[a4paper,11pt]{article}
\usepackage{xcolor}
\usepackage[utf8]{inputenc}
\usepackage{graphicx}
\usepackage{caption}
\usepackage{amsmath}
\usepackage{hyperref}

\title{Molecular Identification via Molecular Fingerprint extraction from Atomic Force Microscopy images}

\author
{Manuel Gonz\'alez-Lastre$^{1}$,  Pablo Pou$^{1,2}$ , Miguel Wiche$^{3,4}$, Daniel Ebeling$^{3,4}$,  \\ Andre Schirmeisen$^{3,4}$, and  Rub\'en P\'erez$^{1,2,\ast}$\\
\\
\normalsize{$^{1}$Departamento de F\'isica Te\'orica de la Materia Condensada,}\\ \normalsize{Universidad Aut\'onoma de Madrid, E-28049 Spain} \\
\normalsize{$^{2}$Condensed Matter Physics Center (IFIMAC),} \\ \normalsize{Universidad Aut\'onoma de Madrid, E-28049 Madrid, Spain} \\ 
\normalsize{$^{3}$Institute of Applied Physics, Justus Liebig University Giessen, Giessen, Germany} \\
\normalsize{$^{4}$Center for Materials Research, Justus Liebig University Giessen, Giessen, Germany}
\\
\normalsize{$^\ast$ ruben.perez@uam.es}
}

\date{\today}

\begin{document}

\begin{@twocolumnfalse}
\maketitle
\begin{abstract}

Non--Contact Atomic Force Microscopy with CO--functionalized metal tips (referred to as HR-AFM) provides access to the internal structure of individual molecules adsorbed on a surface with totally unprecedented resolution. Previous works have shown that deep learning (DL) models can retrieve the chemical and structural information encoded in a 3D stack of constant-height HR--AFM images, leading to molecular identification. In this work,  we overcome their limitations by using a well-established description of the molecular structure in terms of topological fingerprints, the 1024--bit Extended Connectivity Chemical Fingerprints of radius 2 (ECFP4), that were developed for substructure and similarity searching. ECFPs provide local structural information of the molecule, each bit correlating with a particular substructure within the molecule. Our DL model is able to extract this optimized structural descriptor from the 3D HR--AFM stacks and use it, through virtual screening, to identify molecules from their predicted ECFP4 with a retrieval accuracy on theoretical images of 95.4\%. Furthermore, this approach,  unlike previous DL models, assigns a confidence score, the Tanimoto similarity, to each of the candidate molecules, thus providing information on the reliability of the identification.
 By construction, the number of times a certain substructure is present in the molecule is lost during the hashing process, necessary to make them useful for machine learning applications.  We show that it is possible to complement the fingerprint-based virtual screening with global information provided by another DL model that predicts from the same HR--AFM stacks the chemical formula, boosting the identification accuracy up to a 97.6\%.  Finally, we perform a limited test with experimental images, obtaining promising results towards the application of this pipeline under real conditions

\end{abstract}
\end{@twocolumnfalse}

\section{Introduction}\label{sec1}

Atomic Force Microscopy (AFM) operated in the frequency modulation (FM) mode in ultra--high vacuum conditions (commonly known as Non--Contact AFM, NCAFM) has become an essential tool for nanoscience~\cite{garcia2002,GiessiblRevModPhys2003}.  NCAFM allows us to explore and manipulate matter at the atomic scale through the interaction between a sharp apex probe and the sample. The functionalization of AFM metal tips with closed-shell molecules, in particular with CO, provides access with totally unprecedented resolution to the inner structure of small organic molecules adsorbed on surfaces~\cite{GrossScience2009,JelinekJPCM2017,GrossAC2018,zhong2020review}. Since the first High--Resolution (HR) AFM  image of the pentacene molecule~\cite{GrossScience2009}, this striking resolution has been exploited  to disclose bond orders~\cite{GrossScience2012}, to image frontier orbitals~\cite{GrossAC2018}  and charge distributions,   and to track the intermediate products of chemical reactions~\cite{deOteyzaScience2013}. Nowadays, HR--AFM has become an essential tool for on-surface chemistry~\cite{deOteyzaScience2013,clair2019controlling} and fundamental catalysis studies~\cite{AltmanACR2015}.

The utmost  resolution provided by HR--AFM arises from the Pauli repulsion between an inert probe like  CO probe with the electronic charge distribution of the sample molecule~\cite{gross2009chemical,Moll2010} modified by the electrostatic interaction between the potential created by the sample and the charge distribution associated with the oxygen lone pair at the probe~\cite{ellner2019molecular,VanDerLitPRL2016,HapalaNatComm2016}. The flexibility of the bond between the CO and the last atom of the metal probe magnifies the saddle lines of the total potential energy surface sensed by the CO, further enhacing the resolution~\cite{HapalaPRB2014}.

This exquisite sensitivity to the sample charge density immediately rises the question whether we can go beyond structure and use HR--AFM as a molecular identification tool. Given the capability of HR--AFM to address individual molecules, such a tool would not only serve to on--surface chemistry applications but has the potential to overcome some of the fundamental limitations of  the spectroscopic techniques~\cite{HanssenACIE2012} such as vibrational spectroscopy (Fourier Transform Infrared (FTIR) and Raman spectroscopies)~\cite{Balan2019Sep}, nuclear magnetic resonance (NMR)~\cite{Simpson2012Nov}, or mass spectrometry~\cite{Meringer2013May, DeVijlder2018Sep} traditionally used for molecular identification.

For molecular identification solely based on HR--AFM, the repulsive nature of the CO-sample interaction prevents the application of force spectroscopy protocols, based on the determination of maximum attractive forces, that achieved single-atom chemical identification~\cite{SugimotoNature2007}.

Attempts to discriminate atoms in molecules by HR--AFM have been based so far either on differences found in the tip-sample interaction decay at the molecular sites~\cite{ellner2019molecular, HeijdenACSNano2016} or on characteristic image features associated with the chemical properties of particular molecular moieties
~\cite{GuoLangmuir2010,SchulerJACS2015,SchulerChemSci2017,ellner2019molecular,EXP,tschakert2020,zahl2021TMA,GrossAC2018,zhong2020review,SchulzApJL21,kaiser22}. For instance, sharper vertices are displayed for substitutional N atoms on hydrocarbon aromatic rings~\cite{GuoLangmuir2010,ellner2019molecular,HeijdenACSNano2016} due to their lone pair. Furthermore, the decay of the CO-sample interaction over those substitutional N atoms is faster than over their neighboring C atoms~\cite{ellner2019molecular,HeijdenACSNano2016}. In general, due to their slower charge density decay, C atoms in aromatic rings are usually sensed as more repulsive than N, which, in turn, is more repulsive than oxygen. Halogen atoms can also be distinguished in AFM images thanks to their oval shape (associated to their $\sigma$-hole) and to the significantly stronger repulsion  compared to atoms like nitrogen or carbon~\cite{tschakert2020}.
Although promising, these rules do not represent a reliable solution to the atom identification problem as the molecular environment plays an important role:  C atoms in carboxylic groups literally disappear from the image of  trimesic acid (TMA) self-assembled networks as they are much less repulsive than the neighboring O atoms in the acid moiety, that strongly attract the electronic charge towards them~\cite{zahl2021TMA}. Furthermore, small height differences can significantly modify the images~\cite{mDBPc}, leading in many cases to contrast inversion with respect to the above rules.

The previous analysis suggests that not a single HR--AFM image, but a 3D stack of constant--height images covering a range of relevant tip heights is needed to provide enough information on the molecular electronic charge distribution to disentangle the contribution of the bonding topology, the chemical composition and the internal corrugation of the molecule to the contrast of the HR-AFM images. While 2D features, like the sharper vertex associated with N atoms~\cite{ellner2019molecular}  are easily recognized by a human via visual inspection, handling 3D information to discriminate, for example,  among the different halogens (that produce the same oval-shape contrast but with different decays moving out of molecule~\cite{tschakert2020}) calls for the application of a Machine Learning (ML) approach.
In particular,  Deep Learning (DL) has proven to be a powerful tool for learning long--range, complex correlations over large sets of images using a data--centric approach. Convolutional Neural Networks (CNNs)~\cite{krizhevsky2012imagenet} have been employed over 3D stacks of constant-height AFM images with remarkable success at different tasks. In 2020, Alldritt et al.~\cite{alldritt2020} developed a CNN model that obtained information about the 3D molecular structure from an 3D image stack by predicting the van der Waals spheres representation of the molecule. They also reported a preliminary test for the prediction of the chemical composition, with modest but promising results. Later in 2022, Oinonen et al.~\cite{Foster2022MRSbulletin} created a pipeline for obtaining the molecular graph also from 3D image stacks. This pipeline consisted of a CNN that extracted a point cloud representation of the atoms, a peak finding algorithm and a combination of Multilayer Perceptron (MLP) and Graph Neural Network (GNN) models to classify each node and assign the bonds. The detection of atomic positions worked quite reliably even for relatively large
molecules such as PTCDA (3,4,9,10-Perylenetetracarboxylic dianhydride CID: 67191)although there were inconsistencies for  non-planar  systems (error rate of $\sim20$~\%) and the model was sensitive to the choice of coordinate system. However, the compositional analysis, that was restricted to families of atoms --1: (H), 2: (C, Si), 3: (N, P), 4: (O, S), 5: (F, Cl, Br)--, showed errors up to 30\%  for the family (N,P), that was commonly mistaken with  the C- and O-groups.

In previous work, we have addressed the problem of complete molecular identification (structure and composition) of quasi planar organic molecules  with no prior information about them using two different DL approaches, taking as input a stack of 10 constant-height HR-AFM images covering the range of tip-sample distances commonly used for AFM imaging, spanning a distance variation of 1~\AA. 
Firstly, we framed it as an image captioning challenge and used multimodal networks~\cite{Carracedo2023_IUPAC} to solve it. Each multimodal network (M-RNN) included a CNN for image analysis and a Recurrent Neural Network (RNN) for  language processing. The first network took as input the 3D image stack and provided the attributes, the IUPAC terms corresponding to all the chemical groups present in the molecule. The second M-RNN exploited both  the  3D image stack and the attributes provided by the first M-RNN to predict the IUPAC name of the molecule, that completely describes the structure and composition of the molecule.
The  determination of the chemical groups within the molecule had a 95\% accuracy, showing that HR--AFM images did carry significant chemical information and that the CNN model is able to retrieve it. For the prediction of the complete IUPAC name, although the model outperforms most applications of RNN to language translation, the accuracy was limited to 76\% using the cumulative 4-gram BLEU metric~\cite{papineni2002bleu}, the standard metric for natural language processing.  This performance drop is probably related to intrinsic limitations of RNNs models and to the IUPAC formulation rules, specifically designed for humans but not particular suitable for machine learning applications. 

In order  to overcome this language limitation, we devised a completely new perspective using visualisation techniques that map images onto images~\cite{Carracedo2024_CGAN}. Our Conditional Generative Adversarial Network (CGAN) converts the image stack  into a ball-and-stick depiction, where balls of different color and size represent the chemical species and sticks represent the bonds, providing, in this way, complete information on the structure and chemical composition. As an additional advantage, this approach  can handle images  containing groups of molecules bonded by hydrogen or halogen-bond interactions or  molecular fragments that cannot be described by the IUPAC formulation. 
To estimate the accuracy of our identification method we used a global assessment  and two specific evaluations focused on either structure or composition.  The CGAN model achieved a remarkable 74\% of perfect predictions, that increased to 95\% (96\%) when considering only structure (composition). Our criteria in the total accuracy and the structure accuracy was really tough as a prediction was considered correct only if there was a perfect match (in all the predictions, most of the structure is revealed correctly, providing valuable information about the molecule, in spite of been considered as incorrect in the determination of the accuracy.)  

\begin{figure*}[b!]
    \centering
        \includegraphics[width = 0.9\linewidth]{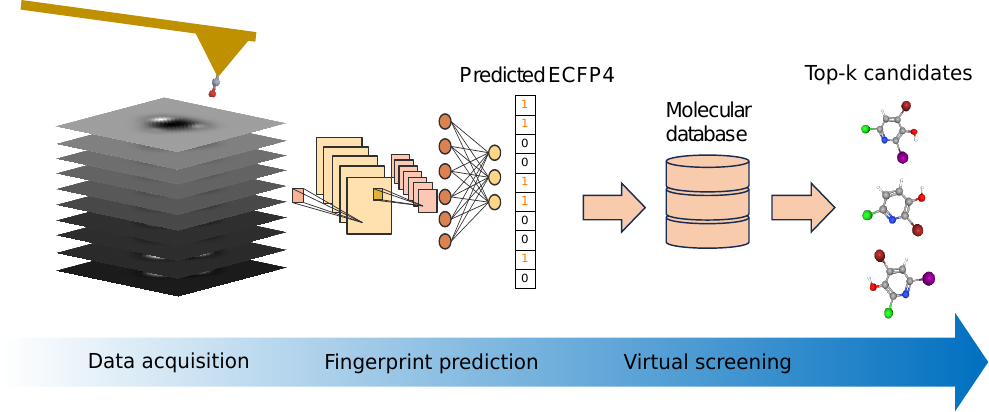}
\caption{Diagram of the molecular identification pipeline. From the experiment, we obtain the 3D HR--AFM stack consisting of 10 constant--height images, which is fed to our neural network to extract the Extended Connectivity Topological Fingerprints (ECFP4). Then, we perform a virtual screening with the predicted fingerprints against a molecular database molecule/fingerprints pairs and rank by decreasing tanimoto similarity.}
\label{fig:inference_pipeline_arrow}
\end{figure*}

The results of the  two DL models described above  show the potential for chemical and structural identification of molecules encoded in HR--AFM images. However they are still limited by the deficiencies of the IUPAC nomenclature as a language in the M-RNN model and by the visual character of the information retrieved by the CGAN, perfectly informative for a human but not useful for its possible use for a prediction of the molecular properties based on the chemical information stored in the HR--AFM images.
Here, we seek to overcome these limitations by using an alternative, well-established description of the molecular structure in terms of topological fingerprints~\cite{Todeschini2000Sep},  that were developed for substructure and similarity searching.
In particular, we have selected a widely used and optimized topological fingerprint, the 1024--bit Extended Connectivity Chemical Fingerprints of radius 2 (ECFP4)~\cite{rogers_hahn_2010}. We show that we can design and train (with the QUAM--AFM database \cite{QUAM-AFM_JCIM}) a DL model that is able to extract this optimized structural descriptor from the 3D HR--AFM stacks and use it, through virtual screening~\cite{Willett2006Dec}, to identify molecules from their predicted ECFP4 with very high accuracy (see Fig.~\ref{fig:inference_pipeline_arrow}).

ECFPs, developed specifically for structure-activity modeling, are circular fingerprints with a number of useful qualities: they can be very rapidly calculated; they are not predefined,  and can represent an essentially infinite number of different molecular substructures  (including stereochemical information). 
ECFPs have proven to be useful in several applications including virtual screening, quantitative structure--activity relationship (QSAR) modeling~\cite{Coley2017Aug,Wu2018Jan} and similarity searching~\cite{Winter2019Feb}. Particularly relevant examples are the use of ECFPs applied to different molecular datasets for  the prediction of electronic properties, solubility and binding affinities for bio-molecular complexes~\cite{Wu2018Jan}, and the recent application to predict compounds with high antibiotic activity and low cytotoxicity~\cite{Wong2024Feb}.

The rest of the paper is organized as follows. After introducing our model for predicting the molecular fingerprints, the 1024--bit ECFP4 ~\cite{rogers_hahn_2010}, of a target molecule and exposing other methodological details used in the work, we show the performance of the model for fingerprint extraction from 3D HR--AFM stacks by using the Tanimoto similarity~\cite{Bajusz2015Dec, Butina1999Jul}. 
As the direct reconstruction of molecular representations from ECFPs is far from being straightforward~\cite{Ucak2023Dec}, we have chosen a virtual screening process as the strategy for molecular identification. Our results show that molecules can be identified from the predicted ECFP4 with very high accuracy (95.4\%). This method, unlike previous works~\cite{Foster2022MRSbulletin, Carracedo2023_IUPAC, Carracedo2024_CGAN}, has the additional advantage of  selecting an arbitrary number of candidate molecules and assigning a confidence score, the Tanimoto similarity~\cite{Bajusz2015Dec, Butina1999Jul}) to each one of them, thus providing information on the reliability of the identification. This approach let us identify the correct molecule even when the prediction of the fingerprint is partially wrong.\\
By construction, ECFPs provide local structural information and the frequency of the identified substructures is lost during the hashing process necessary to map them into a fixed sized vector (see Methods). To address this limitation, we complement the fingerprint-based virtual screening with global information from another deep learning model that predicts the chemical formula from the same high-resolution atomic force microscopy (HR-AFM) stacks, enhancing the identification accuracy to 97.6\%. Finally, we conducted a limited test with experimental images, yielding promising results that support the feasibility of applying this pipeline under real-world conditions.

\section{Methods}\label{sec11}

\subsection{Molecular fingerprints}
\label{section:ECFP4}

Molecular fingerprints~\cite{Todeschini2000Sep} are  representations of the chemical structure of the molecule optimized for substructure searching and machine learning tasks. In molecular fingerprints, each integer represents the presence of a particular substructure. %
In our work, we have chosen the Extended Connectivity Fingerprints (ECFPs)~\cite{rogers_hahn_2010}, a class of topological fingerprints that can be efficiently computed and represent an essentially infinite number of different molecular substructures.

The ECFP generation process begins with the assignment of an initial integer identifier to each atom in the molecule. These identifiers are typically based on atom types and incorporate properties such as the valence, atomic mass and so forth. 
Following this, an iterative neighborhood expansion process takes place up to a defined radius. In each iteration, a new integer identifier is created for each atom by hashing its current identifier together with those of its immediate neighbors, in order to incorporate information from the atom's local environment. 
In ECFP4, the radius is set to second neighbors. Finally, we map the fingerprints to a fixed--size 1024--bit vector. To obtain the index of the ``on'' bits in the final bit vector, we use the modulo operator on each integer. This hashing step, although necessary for machine learning applications~\cite{rogers_hahn_2010,Wu2018Jan}, produces a loss of information: 
firstly, the frequency of each substructure's occurrence within the molecule is lost; secondly, different integers can be mapped to the same index (a situation referred to as a ``bit collision''~\cite{rogers_hahn_2010}).

In our implementation, we have used the RDKit~\cite{landrum} library to compute the molecular fingerprints from the SMILES code of the molecules, obtained from the PubChem~\cite{Pubchem} repository.

\subsection{Tanimoto similarity and virtual screening} \label{sec:virtual_screening}
Labelling molecules with these fingerprints allows an easy and fast quantification of the difference or similarity between two molecules A and B. We have chosen the Taminoto Similarity~\cite{Bajusz2015Dec, Butina1999Jul}, $S_{A,B}$, calculated as:
\begin{equation}
    0 \leq S_{A,B} = \frac{c}{a+b-c} \leq 1
\end{equation}
where $a$ is the number of on bits in molecule A, $b$ the number of on bits in molecule B and $c$ the number of bits that are on in both molecules~\cite{Bajusz2015Dec}. The closer the value is to 1, the more similar molecules A and B are. Therefore, $S_{A,B}=1$ means A and B are the same, except for the limitations due to the local character of the fingerprints and the information lost in the hashing step of the fingerprint generation process. 

Using this similarity metric, we can identify and rank candidate molecules via virtual screening as described in~\cite{Willett2006Dec}: firstly, the Tanimoto similarity~\cite{Bajusz2015Dec, Butina1999Jul} between the predicted fingerprint and each molecule in the database is computed. Then, the candidates are ranked by decreasing order. At the end, the top-$k$ candidates are returned as the output of the screening process, where $k$ is an optional parameter set by the user. Here, the Tanimoto similarity serves both as a ranking metric and as the model's confidence in the prediction.

\subsection{Architecture of models} \label{subsection: model_architecture}

In this work, we have developed two CNN models: (i) a multilabel classification model for the prediction of molecular fingerprints; and (ii) a regression model for the count of each atomic species within the structure, from which we construct the chemical formula. 

The molecular fingerprint model is an adaptation of EfficientNet-B0~\cite{efficientnet}, where we change the first convolutional layer from 3 to 10 channels so it accepts stacks of 10 constant-height HR-AFM images as input, allowing the model to take the whole $z$--range at once. The final layer consists of a Dense layer of size 1024 with sigmoid activation. A critical step for improving the model's performance on experimental images was to substitute the first BatchNorm layer of the EfficientNet model for a Dropout layer with dropout probability $p=0.5$. The dropout layer prevents co-adaptation of neurons~\cite{srivastava2014dropout}, what makes the model robust against experimental conditions (noise, plane tilting, etc.) which are not present in the simulated images used for the training (see below).
The chemical formula model is constructed in the same way, but using as the final layer a Dense layer of 10 neurons with ReLu activation.

\subsection{Training and evaluation}

This work aims to create an end--to--end molecular identification tool that uses a 3D stack of  experimental AFM images as input. However, training neural networks requires a high amount of labeled samples and there is currently no such dataset for experimental images. For this reason, we train and evaluate our models primarily on simulated images from the QUAM--AFM~\cite{QUAM-AFM_JCIM} dataset, containing 165 million HR-AFM images theoretically generated from a selection of 685,513 isolated quasi-planar molecules from PubChem~\cite{Pubchem} that span the most relevant bonding structures and chemical species in organic chemistry. 
The QUAM--AFM can be freely downloaded~\cite{QUAM-AFM_repository_CM}.
For each molecule, HR-AFM images were simulated for 10 tip--sample distances considering six different values of cantilever oscillation amplitude and four values of the tilt stiffness of the CO molecule to cover a wide range of experimental operation conditions.

We performed data curation by removing the non--live molecules~\cite{Kim2016Sep} from our dataset. These molecules were accessible through the PubChem API at the date of creation of QUAM--AFM, but have since then become inaccessible. After this step, our train/validation/test split consisted of 285k/15k/280k
randomly sampled molecules respectively. As the training set is so huge, we divide each epoch in 10 virtual epochs and compute validation metrics at the end of each of these virtual epochs (see Supplementary Information, Fig. S1).

For the molecular fingerprint model, the binary cross-entropy with logits loss function, equipped with balanced positive weights, was used as the training criterion:

\begin{align}
L = -\frac{1}{N} \sum_{i=1}^{N} \bigg[ &p_c \cdot y_i \cdot \log(\sigma(x_i)) + (1 - y_i) \cdot \log(1 - \sigma(x_i)) \bigg],
\end{align}
where \( y_i \) is the ground truth for the \( i \)-th bit of the fingerprint, \( \sigma \) is the sigmoid function and \( \sigma(x_i) \) the probability predicted by the model for that same bit.
The \( p_c \) parameter is used to give more weight to correctly predicting the on bits (see section~S3). Molecular fingerprints are quite sparse and without this term, the network could be trained into only predicting 0's.\\

Regarding the training strategy, we initialized the fingerprint model from pre-trained weights~\cite{efficientnet}. The bias of the last layer is initialized with a prior to accelerate the convergence of the model (see section~S1 for details). Then, the model was trained until  a plateau at mean Tanimoto similarity $S=0.88$ is reached in the validation set (Figure~S1a). We select the last checkpoint as our models' weight.

For the model that predicts the chemical formula, we followed a transfer learning strategy: we cloned the weights from the backbone of the fingerprint model to the backbone of the chemical formula model and trained it end-to-end (Figure~S1b). Since we had a very good pre-training, the model is almost converged on the first virtual epoch (MSE: $0.1 \, \text{atoms}^2$), after which our validation oscillates for the rest of the training.
The training hyperparameters of both models were chosen to be the standard for classification and regression tasks (Table~S1). For the chemical formula model, we define accuracy as the probability of perfect  prediction, a very hard metric on our model as a miscounting in the number of hydrogens is considered a failed prediction. 

It is important to stress that,  during the training of both models, strong data augmentations were applied in order to regularize the model and to reproduce the effect of experimental conditions on the HR--AFM images. These augmentations include rotations, translations, shears, in/out zooms, and gaussian noise (see Table~S2 for details).

\subsection{DFT calculations and simulation of HR--AFM images}\label{sec:dft_ptcda} 

The HR--AFM images in the QUAM--AFM dataset were simulated using the  gas phase molecular structure available from PubChem. However, when molecules are deposited in a substrate, the interaction with the surface changes the molecule corrugation, which translates into differences in the contrast of the images.

To study the ability of the model to identify molecules in a substrate, we simulated the adsorption of PTCDA  
on Cu(111) and Ag(111) surfaces. For the Cu(111) slab, we used a unit cell of size 20.4 x 22.08 x 30.83 \AA{}  (including about 25.49 \AA{} of vacuum). The slab model contained 3 layers of copper, making 238 atoms in total. As starting geometry, we placed the PTCDA molecule in its gas phase structure at 2.86 \AA\ above the slab.
For the geometry relaxation, DFT calculations were carried out using the VASP package~\cite{kresse96} with a cutoff energy of 425 eV for the plane-wave basis set. The projector augmented wave method~\cite{bloch1994,kresse1999} was used to build the pseudopotentials of all constituent species. We use the PBE generalized gradient
approximation~\cite{PBE1996} for the exchange-correlation part of the energy and the semiempirical DFT-D3 dispersion correction~\cite{grimme2010a} to model  the Van der Waals interaction. The PTCDA structure adsorbed on Cu(111) was converged using a conjugate gradient algorithm until forces upon atoms were smaller than 0.01 \text{eV/\AA} while each electronic self-consistent loop was calculated with a precision of $10^{-5}~\text{eV}$. A vertical vacuum region of 22.4 \AA{} was established between the periodical images and a dipole correction along the z-axis was also used. As we were only interested in the effect of the substrate on the molecular corrugation, we left the substrate fixed and sampled the Brillouin zone using only the \(\Gamma\) point.

The geometry of the PTCDA molecule adsorbed on Ag(111) was obtained from ref~\cite{tesis_emiliano}, which was calculated with the same parameters but an energy cutoff of 400 eV and convergence criterium of $10^{-6}~\text{eV}$ for the SCF calculation.

HR--AFM images are simulated with the same model used to generate the QUAM--AFM data set~\cite{QUAM-AFM_JCIM}: an approximate implementation of the full density based model (FDBM)~\cite{ellner2019molecular}  in the latest version of the PPMAFM code~\cite{liebig2020quantifying,QUAM-AFM_JCIM}. Only the molecular structures obtained from the adsorption calculations are included in the corresponding HR--AFM simulations.

\section{Results and discussion}\label{sec2}
\subsection{Predicting ECFP4 from HR--AFM images.}

\begin{figure}[t!]
    \centering
    \includegraphics[width = 0.6\linewidth]{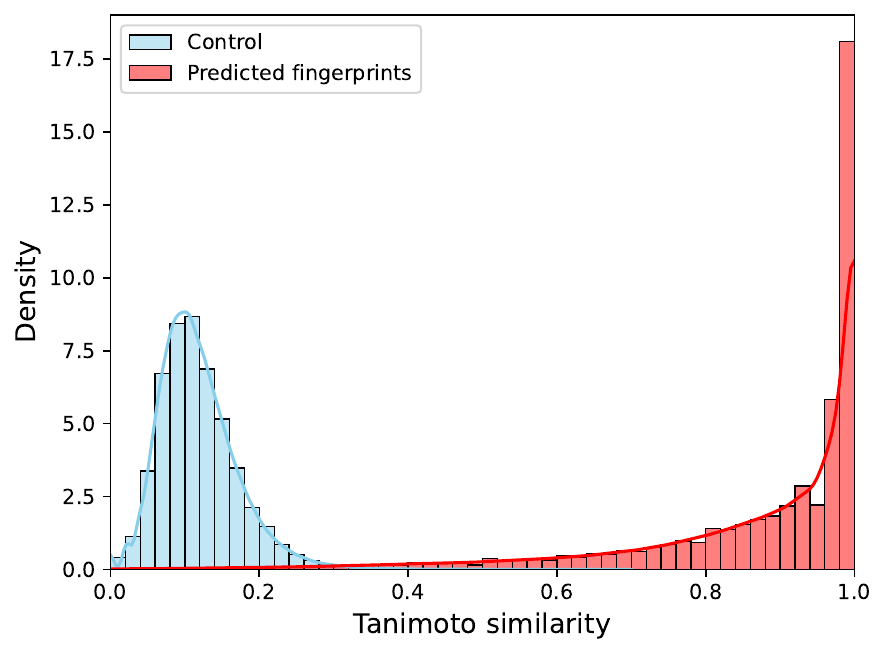}
    \caption{Tanimoto similarity distributions between predicted and ground truth fingerprints (red) and randomly drawn molecules fingerprints (blue). We can extract enough chemical information to distinguish a molecule from the bulk.} 
    \label{fig:tanimoto_distribution}
\end{figure}

First, we evaluate the performance of our chemical fingerprint prediction model.  We use our model to predict the chemical fingerprints from the  3D stacks of HR--AFM images for the 279905 molecules in the test set and plot the histogram of the Tanimoto similarity (red histogram in Fig.~\ref{fig:tanimoto_distribution}) between the predicted and ground truth fingerprints. 
This histogram has a median Tanimoto similarity equal to $0.95$, demonstrating that we can predict the chemical fingerprints from HR--AFM image stacks very accurately. To compare with a baseline, we also compute a control histogram (blue histogram on Fig.~\ref{fig:tanimoto_distribution}) corresponding to the Tanimoto similarity of pairs of randomly chosen molecules. This control histogram has median 0.11 and a very low density for values of Tanimoto greater than $0.4$. From Fig.~\ref{fig:tanimoto_distribution} we can conclude that, in most cases, we won't be able to predict the molecular fingerprints perfectly (S=1). Nevertheless, a not-perfectly predicted still can store enough information of the molecule to be identified. Statistically, a prediction with Tanimoto similarity higher than $0.5$ should be enough to identify a molecule from the HR--AFM images.\\  

The chemical information provided in the predicted fingerprint outperforms previous models~\cite{alldritt2020,Foster2022MRSbulletin,Carracedo2023_IUPAC,Carracedo2024_CGAN}. 
Alldrit et al.~\cite{alldritt2020} focused on structural elucidation and only presented few preliminary results for chemical recognition. Later work~\cite{Foster2022MRSbulletin} addressed molecular identification using GNNs but showed modest accuracy compared to this work. In our work with M-RNNs~\cite{Carracedo2023_IUPAC}, we presented a model that was able to identify the chemical groups in a molecule from a stack of simulated HR-AFM images with 95\% precision, comparable to the present work. However, the retrieved ECFP4s contain more information as they encode both molecular moieties and structural information.  Finally, in our recent CGAN model for balll-and-stick prediction~\cite{Carracedo2024_CGAN}, although remarkable in its prediction for either the structure or composition ($\sim$95\%), the combined performance dropped to $\sim$~76\%.  Furthermore,  the visual character of the information retrieved by the CGAN is not suitable for its subsequent use for the prediction of other molecular properties, as  structure--activity relationships  and similarity searching where the ECFPs have already shown their potential. Thus, the outstanding precision demonstrated by our novel model in forecasting ECFP4 marks a significant advancement in HR-AFM image analysis.

\subsection{Molecular Identification via virtual screening}

One of the main goals of this work is to automate molecular identification.  Our hypothesis is that a molecule can be identified through its predicted ECFP4 performing a virtual screening: we calculate the Tanimoto similarity of the predicted fingerprint against the fingerprints of all the molecules from the reference dataset and retrieve the top candidates, i.e., those with the highest Tanimoto.

\begin{figure}[b!]
    \centering
    \includegraphics[width = 0.6\linewidth]{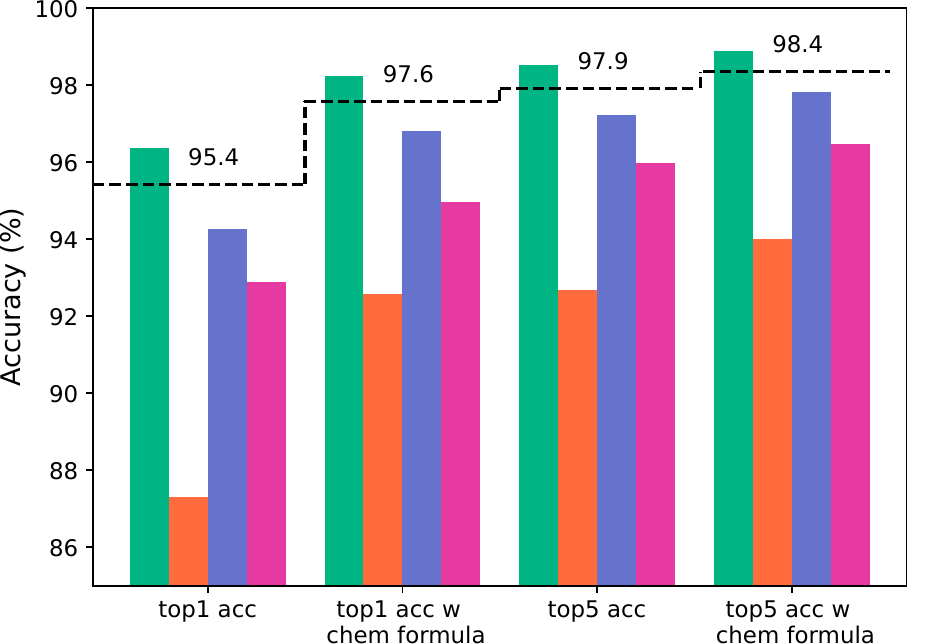}
    \caption{Identification accuracy versus corrugation. We compute the accuracy for molecules with corrugation $<$ 25 pm (green), 25--75 pm (orange) 75--125 pm (purple) and $>$ 125 pm (magenta). Dashed black lines represent the accuracy over all corrugation groups. Enriching the virtual screening with the chemical formula improves accuracy across all corrugation groups.}
    \label{fig:top_1acc_chemical_formula_corrugation} 
\end{figure}

\begin{figure*}[b!]
    \centering
    \includegraphics[width = 0.9\linewidth]{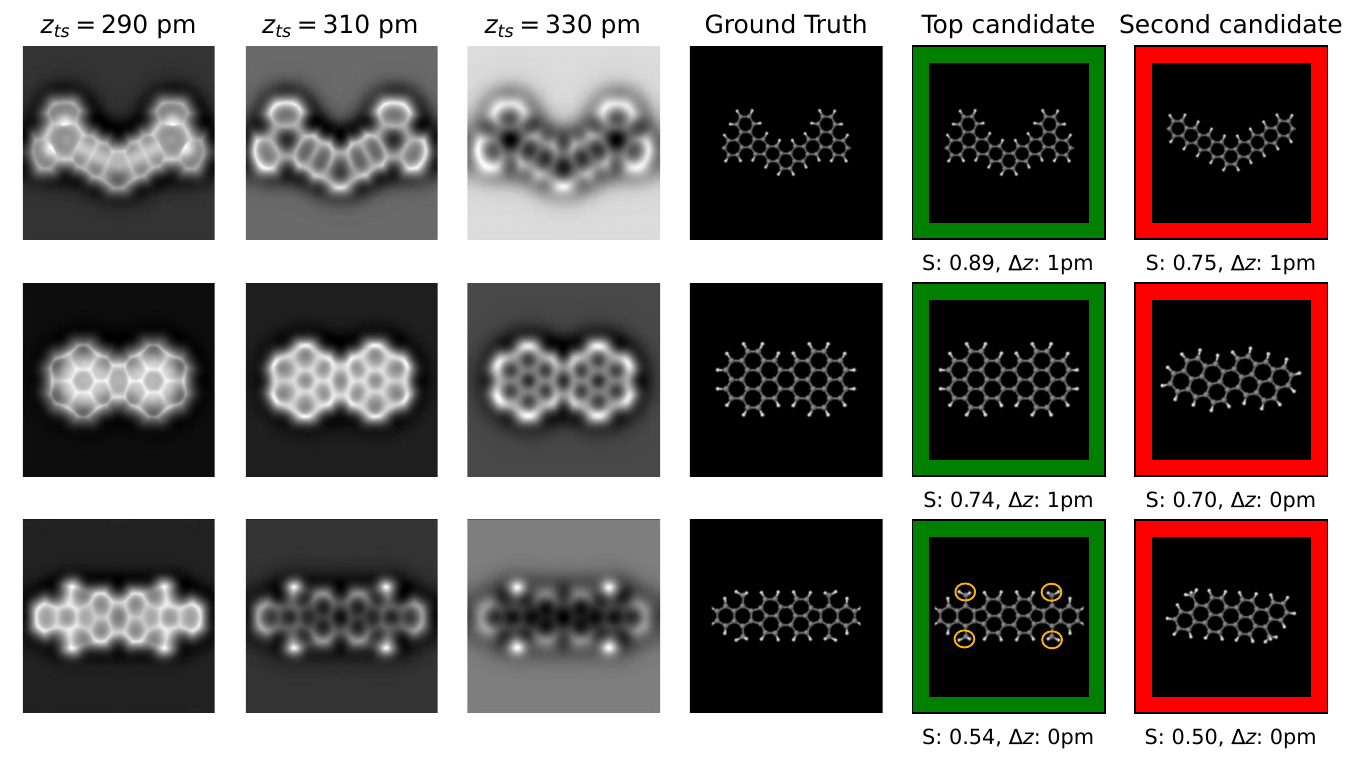}
    \caption{Examples of identification of polycyclic aromatic hydrocarbons over theoretical 3D stacks. Columns from left to right, constant-height AFM images (1-3), ground truth molecule (4) and top (5) and second (6) candidates. Under each candidate, tanimoto similarity, $S$ and corrugation, $\Delta z$ is expressed. Molecules from first to last row are Tetrabenzo(a,c,g,s)heptaphene (CID:143932), Benzo[1,2,3-bc:4,5,6-b'c']dicoronene (CID:636081) and Tetramethyl-Undecacyclo-Tetraconta-Icosaene (CID: 59721948), where methyl groups have been highlighted. The model identifies the correct molecules with high confidence.\label{fig:pah_identification}}
\end{figure*}

\begin{figure*}[t!]
    \centering
    \includegraphics[width = 0.9\linewidth]{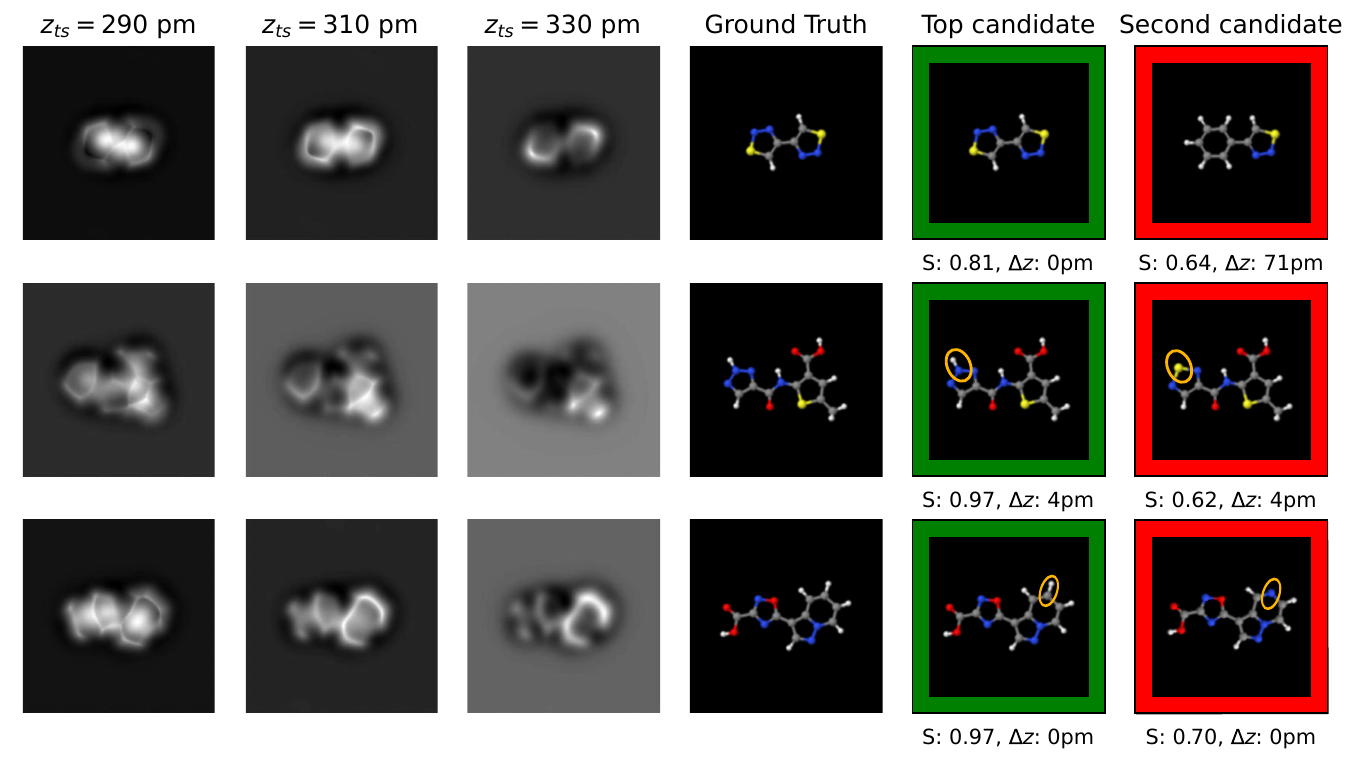}
    \caption{Examples of identification of molecules with nitrogen, oxygen and sulfur atoms. From first to last row: 4,4'-Bi[1,2,3-thiadiazole] (CID:2748722), 5-methyl-2-(2H-triazole-4-carbonylamino)thiophene-3-carboxylic acid (CID:63616469) and 5-Pyrazolo[1,5-a]pyridin-3-yl-1,2,4-oxadiazole-3-carboxylic acid (CID:103122053). IN the last two rows, the differences between candidate molecules have been highlighted to guide the reader.\label{fig:chemical_speciesl}}
\end{figure*}

\begin{figure*}[b!]
    \centering
    \includegraphics[width = 0.9\linewidth]{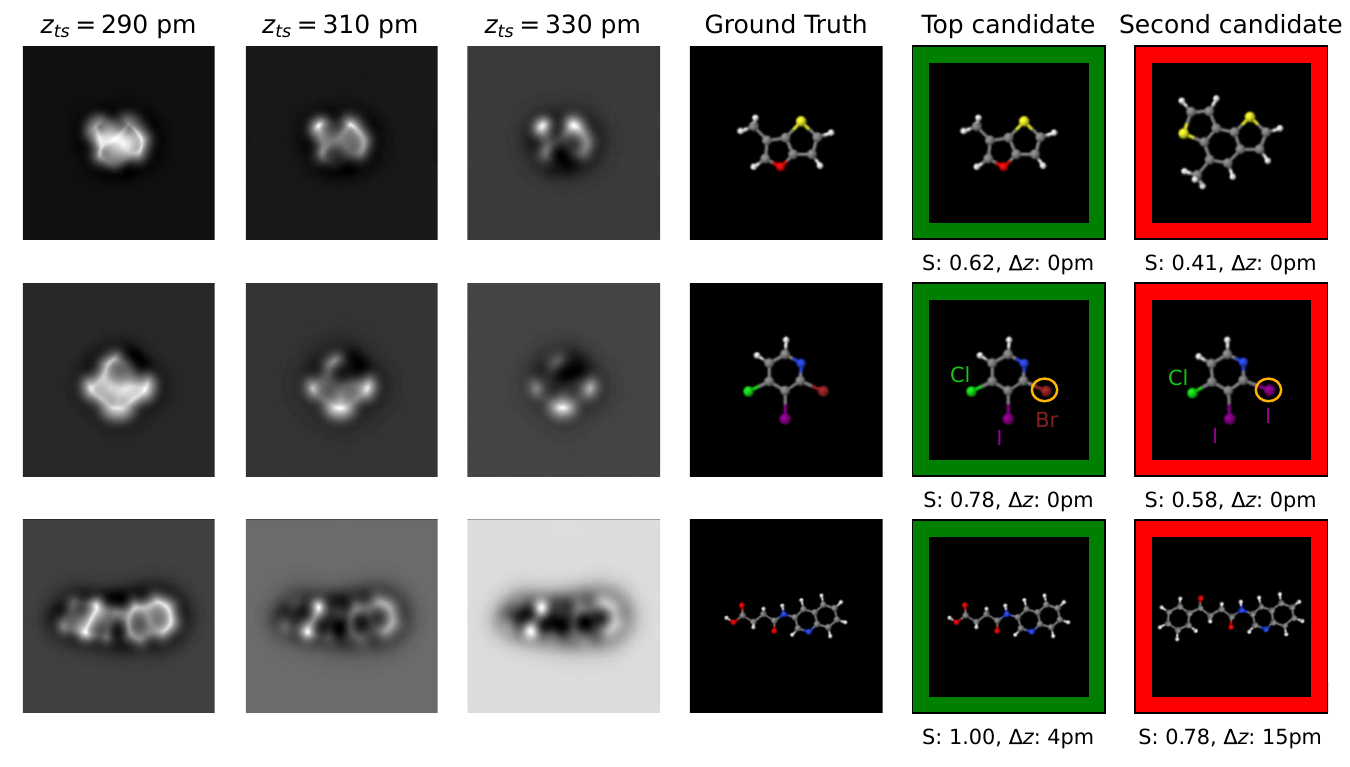}
    \caption{Examples of identification of molecules with chemical species of the same group. Columns organized as in Figure~\ref{fig:pah_identification}.  From first to last row: 3-Methylthieno[3,2-b]furan (CID:58899415), 2-Bromo-4-chloro-3-iodopyridine (CID: 59332995) and 4-Oxo-4-(quinolin-3-ylamino)butanoic acid (CID: 861757).\label{fig:identification_3_cases}}
\end{figure*}

\begin{figure*}[b!]
    \centering
    \includegraphics[width = 0.9\linewidth]{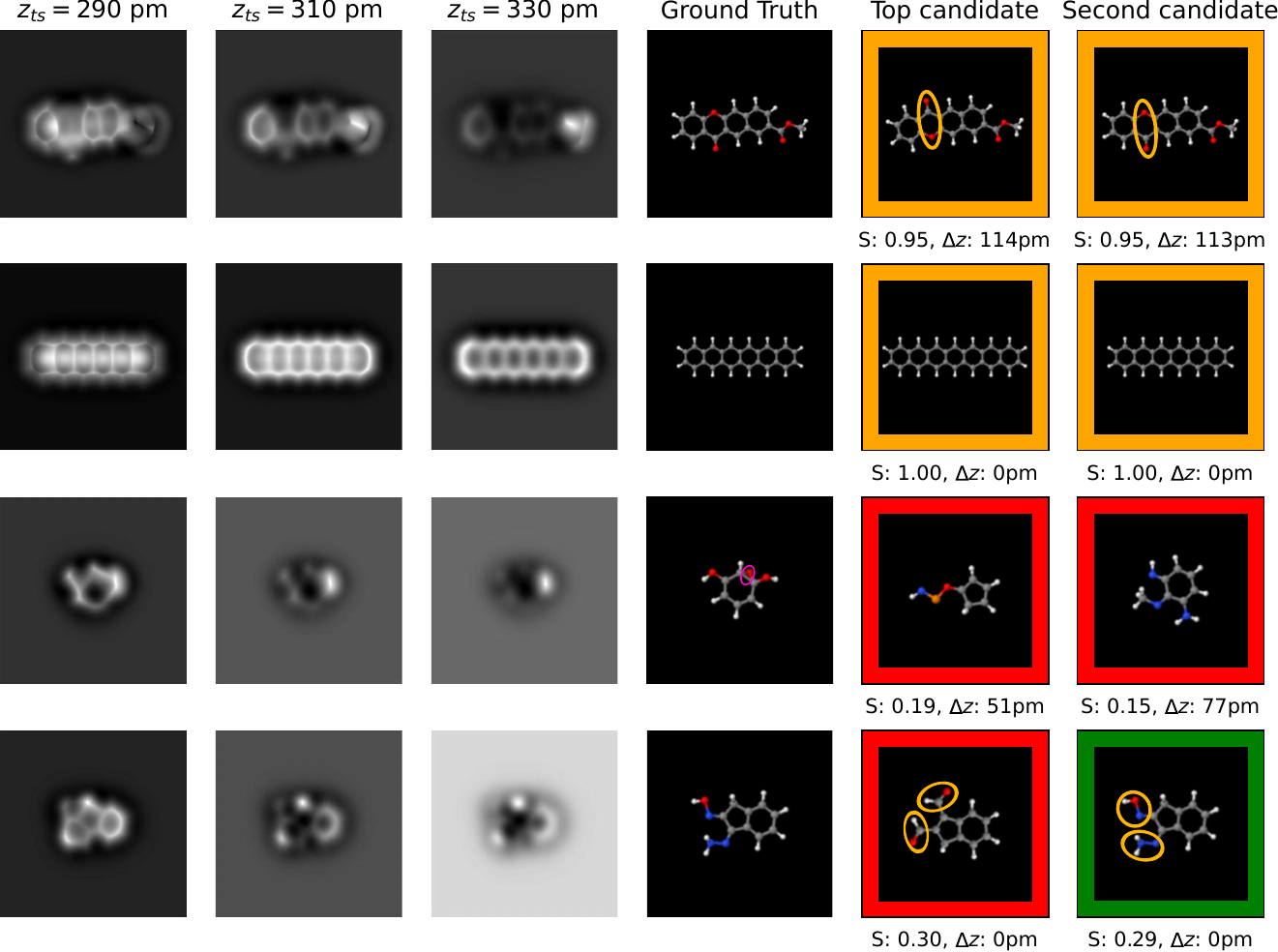}
    \caption{Examples of incorrect identifications. Molecules from first to last row are Methyl 12-oxobenzo[b]xanthene-9-carboxylate (CID:135178930), Hexacene, (CID:123044), Resorcinoxide (CID:129866873) and N-(3-hydrazinylidene-1H-inden-2-ylidene)hydroxylamine (CID:137221883). In Resorcinoxide, there is an oxygen atom inaccessible to the tip, as it is under the benzene ring (highlighted in purple). The identification fails not because of the model, but rather because the AFM doesn't have access to this region of the molecule.}
    \label{fig:vs_bad_pred}
\end{figure*}

On our test-dataset of simulated images of 279905 molecules, we achieve a top1 and top5 retrieval accuracy of 95.43\% and 97.92\% respectively (see Fig.~\ref{fig:top_1acc_chemical_formula_corrugation}). This means that our model is able to correctly identify the molecule in practically all the cases of our dataset which includes a large variety of homo- and hetero- acyclic or cyclic compounds with the most relevant functional groups including alkanes, alkenes, alkyne, alcohols, thiols, ethers, aldehydes and ketones, carboxylic acids, amines, amides, imines, esters, nitriles, nitro and azo compounds, halocarbons, and  acylhalide.  Figures~\ref{fig:pah_identification},~\ref{fig:chemical_speciesl},~\ref{fig:identification_3_cases} show a few examples of correct identification on different sets of molecules.

Our first test is on polyaromatic hydrocarbons (PAHs) (Fig.~\ref{fig:pah_identification}), which include only two chemical species (carbon and hydrogen). Independently of the number of rings or the bond order distribution, the model correctly identifies these molecules. All these PAHs have correctly been identified, even if the predicted fingerprints weren't 100\% correct (Tanimoto $S=1$). This highlights the robustness of the virtual screening, where even partially correct predictions of the fingerprints is enough to correctly identify the molecule (Fig.~\ref{fig:tanimoto_distribution}). In the first row, the top and second candidates have quite different topologies and the difference in tanimoto similarity is $\Delta S=0.14$. In comparison, the second and third rows candidates with very similar topologies, which is why the difference in tanimoto is less than a third ($\Delta S = 0.04$ in both cases). Since the topologies are so similar, the model is less confident in predicting if the correct molecule is one or the other (even if in the end recognizes the correct structure). 

The molecule in the last row, Tetramethyl-Undecacyclo-Tetraconta-Icosaene (CID: 59721948), shows that the model not only works for perfectly flat cases, but it correctly identifies the presence of methyl groups. The tanimoto ($S=0.54$) is relatively low, comparing with the two previous examples, meaning that the corrugation had an effect on the accuracy of the model.

Next, we consider molecules including nitrogen, oxygen and sulfur atoms besides carbon and hydrogen. The three chosen molecules (see Fig.~\ref{fig:chemical_speciesl}) display a variety of  bonding configurations and include a number of different chemical groups: carboxylic, methyl, and amide groups, thiophene, thiadiazole or oxadiazole rings. Despite the presence of several atomic species and the structural complexity, our model correctly identified the target molecule with high confidence. In the first row, the model correctly predicts the two thiadiazole groups. In the second row, the top two candidates only differ by a triazole vs a thiodiazole (this means a N-H vs a S atom),  while in the third row, the model correctly discriminates between a Pyrazolo[1,5-a]pyridine and  a Pyrazolopyrimidine (a C for an N).

Lastly, we test whether we can distinguish between different bonding coordinations for the same chemical species or among elements in the same chemical family (same column in the Periodic Table). Fig.~\ref{fig:identification_3_cases} shows (first row) how the model is able to recognize and discriminate the furan and thieno groups in the molecule 3-Methylthieno[3,2-b]furan (CID:58899415). We can also discern different halogens (Fig.~\ref{fig:identification_3_cases}, second row) as seen for the 2-Bromo-4-chloro-3-iodopyridine (CID: 59332995) molecule. Finally, in the third row, we see how, for the 4-Oxo-4-(quinolin-3-ylamino)butanoic acid (CID: 861757) molecule formed by a chain with methylene, amide and carboxyl groups, the ECFP4 was perfectly predicted.

As the interaction is so sensitive to the tip--sample distance, the internal corrugation of the molecule is one of the key contributors to the HR--AFM contrast. Disentangling this effect from the bonding configuration and the chemical composition to achieve molecular identification is a major challenge. In our study, we have restricted ourselves to molecules with corrugations smaller $185\,\text{pm}$, that include the presence of methyl groups and are within the height range from where information can be retrieved with the common constant--height operation mode for HR--AFM~\cite{alldritt2020}. 
This limitation arises from both the strength of the Pauli repulsion on the higher atoms and the deflection of the CO probe, that contributes to sharpen the features associated to the higher atoms, but, at the same time, veils  the access to the lower ones, effectively creating regions that are inaccessible to the tip. Figure~\ref{fig:top_1acc_chemical_formula_corrugation} plots the accuracy of the model for molecular identification for molecules from our test set 
of 279905 molecules with corrugations in four different ranges. We do see a drop in accuracy when we move to larger corrugations, but the reduction is rather small ($\simeq$3.5\% for the group with larger corrugation).

Our fingerprint--based identification pipeline has very few misidentifications (less than 5\% of the cases). In Figure~\ref{fig:vs_bad_pred}, we explore four typical failures that help us unveil the limitations of our model and illustrates how some of them can be easily fixed.

The case shown in the first row of Fig.~\ref{fig:vs_bad_pred} is archetypical:  the retrieval fails because the molecular fingerprints of the ground truth molecule (Methyl 12-oxobenzo[b]xanthene-9-carboxylate, CID:135178930) and the top candidate (Methyl 12-oxobenzo[b]xanthene-8-carboxylate, CID:135178929) are indeed the same. Since the radius used to create the fingerprints is limited to next nearest neighbors (the two closest atoms), they cannot capture the switching in the relative position of radicals that are far away. In this case, there is a tie in the Tanimoto similarity and the order of the candidates is arbitrary.

The second row of Fig.~\ref{fig:vs_bad_pred} illustrates another case where molecular identification is hampered by the local character of the fingerprints:  the difference between the predicted Naphthacene (CID:7080) and the second candidate (and, in this case, ground truth) Hexacene (CID:123044) molecules is the number of benzene rings. Our fingerprints are binary, which means that they represent the presence or absence of certain molecular substructures, but they don't retain information about the number of times they are present. This information is lost in the hashing step, subsection~\ref{section:ECFP4}, in the construction of the ECFP4 fingreprints. In these cases, different molecules can have the same ECFP4 while having different chemical formula. The retrieval failure cannot be attributed to the performance of the CNN to extract the chemical information from the HR-AFM image stack but to the fingerprint codification.

On the third row we present a failure with a completely different origin. The Resorcinoxide (CID:129866873) molecule is corrugated ($175\,pm$). In the configuration we have used to calculate the HR-AFM images, an oxygen atom is under the benzene ring, inaccessible to the tip. As the AFM tip is not able to sense the full structure, the HR-AFM images cannot provide enough chemical information, and the prediction of the network fails. In this case, the accuracy is not limited by the model or the choice of molecular descriptor, but the intrinsic limitation of the HR--AFM operated on the constant-height mode to retrieve information for complex 3D structures from a single adsorption configuration.

Finally, in the last row, we show a case where the model has problems extracting the fingerprints from the HR--AFM image stack, as shown by the low Tanimoto similarity of the two top candidates. In particular, the model did recognize the hexagonal and pentagonal rings and the presence of an OH group, but failed to identify the nitrogen atoms in the ground-truth molecule N-(3-hydrazinylidene-1H-inden-2-ylidene)hydroxylamine, CID:137221883) and predicted a molecule with carbons and oxygens instead (3H-indene-1,2-dicarbaldehyde, CID: 129814712) as the top candidate. However, the second candidate, with a very similar Tanimoto similarity, is the ground truth.
Given the success of the model with other molecules containing N atoms, we attribute the failure in this case to the fact that the presence of OH and NH2 groups linked through an additional N atom is quite rare in organic compounds, and, in particular, in our training set. 

The limitations posed by the hashing step (due to the associated information loss) can be solved with an additional model trained to predict the chemical formula from the HR--AFM stack. The accuracy of this model is near perfect (above 99.5\%, Table~S3), except for phosphorus atoms (78.5\%), which are underrepresented compared to the rest of the chemical species in the dataset. This additional model immediately solves the misidentification between  Naphthacene (7 rings)  and Hexacene (6 rings) (Fig.~\ref{fig:vs_bad_pred},  second row).
Thus, the final pipeline for molecular identification consists of a virtual screening using the predicted ECFP4, which outputs k--candidates with decreasing Tanimoto similarity, and a posterior re--ranking of the candidates by calculating the mean squared error of the predicted and ground truth chemical formula.  With this strategy, that combines  local (fingerprint) and global (chemical formula) features,  the identification accuracy  jumps from 95.43\% to 97.59\%, almost reducing misidentifications by half.

\subsection{Effect of the adsorption-induced molecular corrugation}\label{sec:corrugation}

\begin{figure*}[t!]
    \centering
    \includegraphics[width = 0.9\linewidth]{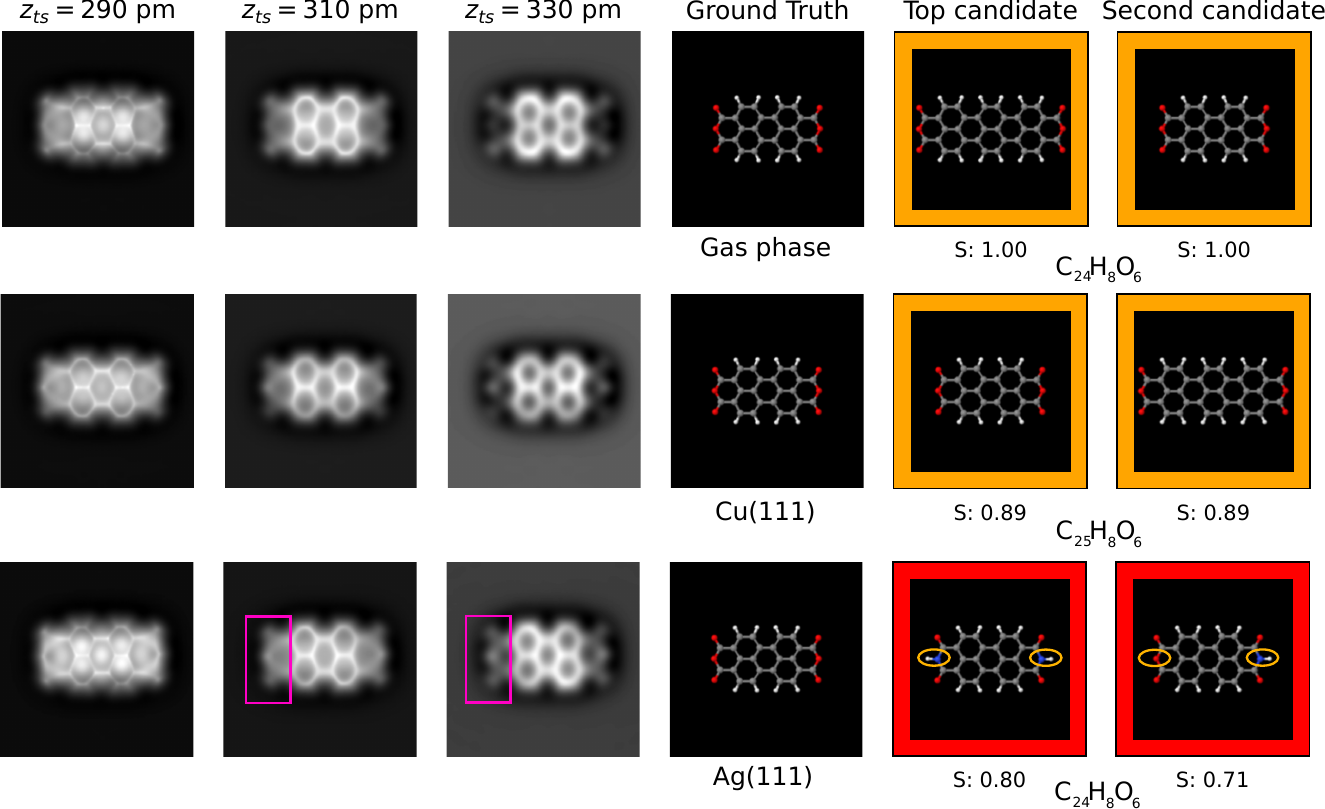}
    \caption{Chemical identification of theoretically generated PTCDA molecules on gas phase (first row) and adsorbed on Cu(111) (second row) and Ag(111) (third row). Tanimoto similarity for each candidate and predicted chemical formula under the candidate images. In gas phase, the fingerprints are predicted perfectly while in Cu(111), the tanimoto drops by 0.1. In Ag(111), the surface pushes away the middle oxygens, increasing their contrast with respect to the gas phase image. The differences in contrast can be clearly seen at  $z_{ts}$ = 310 and 330 pm (purple). The model interprets this contrast as NH groups (blue) and predict the Perylimid molecule as first candidate instead of the PTCDA. 
     \label{fig:ptcda_corrugation_gas_ag_cu}}  
\end{figure*}

Our final goal is to develop a model capable of retrieving the molecular fingerprints from experimental HR--AFM images. In experiments, the molecules are necessarily adsorbed on a substrate, and, due to the molecule-substrate interaction, the adsorption configuration will differ from their gas phase structure. As the data set used for the training of the model is based on HR--AFM images calculated for the gas--phase configuration, it is important to test the ability of the model to identify a molecule from images corresponding to  their structure upon adsorption on different substrates or on different configurations within the same substrate. 

We have addressed this question with the PTCDA molecule, considering HR-AFM images simulated for its gas phase structure and for the adsorption configurations on  both Cu(111) and Ag(111) surfaces, as determined from DFT calculations (see section~\ref{sec:dft_ptcda}  for details). 
Fig.~\ref{fig:ptcda_corrugation_gas_ag_cu} displays some of the simulated images in the HR--AFM 3D stack for the three cases and  the predictions of the model over PTCDA on gas phase (first row), adsorbed on Cu(111) (second row) and Ag(111) (third row). In the gas phase, PTCDA has a perfectly planar geometry and the model achieves a perfect prediction of the molecular fingerprints, with a Tanimoto similarity S=1.  As shown there, virtual screening produces a tie between two structures with the same fingerprints, but the addition of the chemical formula model, that accurately retrieves the chemical composition  $\textnormal{C}_{24}\textnormal{H}_{8}\textnormal{O}_{6}$ from the 3D stack, leads to the proper identification. 

In the case of Cu(111), the interaction with the substrate corrugates the PTCDA structure, pulling the oxygen atoms in the corner towards the surface by 11 pm with respect to the central carbon ring, while the central oxygen is pushed 6 pm above (Fig.~S3).  This corresponds well with the contrast of our simulated HR-AFM image stacks, where the middle oxygen is brighter in the case of Cu(111)--adsorbed molecule than the rest of the O atoms and also brighter than in the images for the gas--phase structure. The outer hexagonal rings are also slightly deformed, with the vortex occupied by the O atom protruding beyond the real O position due to its lone pair~\cite{zahl2021TMA}, as we have also observed in the case of substitutional nitrogen atoms~\cite{ellner2019molecular}. Our model extracts quite accurately the fingerprints (Tanimoto similarity $S= 0.89$) while the  chemical formula  predicts  $ \textnormal{C}_{25}\textnormal{H}_{8}\textnormal{O}_{6}$, not perfect, but good enough to achieve an  unambiguous identification.

For PTCDA on Ag(111), 
the central O atom is pushed up  by 5 pm (Fig.~S3), making them brighter than in the images for the gas--phase structure. The fingerprint model retrieves two molecules with a high Tanimoto similarity, Perylimid ($S=0.80$) --where the central atoms are replaced by NH groups-- as the first candidate, and 1H-2-Benzopyrano[6',5',4':10,5,6]anthra[2,1,9-def]isoquinoline-1,3,8,10(9H)-tetrone (CID: 118580) ($S=0.71$) as the second one, while the chemical formula model predicts the correct composition. This is a tough case, where it is difficult to disentangle the effect of corrugation and chemical composition. The simulated HR--AFM images for Perylimid (some of them are shown in Fig.~S2) are very similar to those calculated for the adsorption configuration of PTCDA on Ag(111), with only subtle differences in the outer areas beyond the O (or N) position. From our experience with other molecular systems~\cite{Carracedo2024_CGAN}, O and NH substitutionals produced very similar charge density distributions and, thus, HR--AFM contrast, slightly more repulsive in the NH case. However, the small upward displacement of the central oxygen results on image features that are very difficult to be discerned from NH groups. 
In summary, this example shows the ability of our identification procedure, combining the fingerprint  and chemical formula models, to cope with the corrugation induced by the molecular adsorption, although further work is needed to assess its accuracy for certain chemical groups.

\subsection{Experimental images}

\begin{figure*}[t!]
    \centering
    \includegraphics[width = 0.9\linewidth]{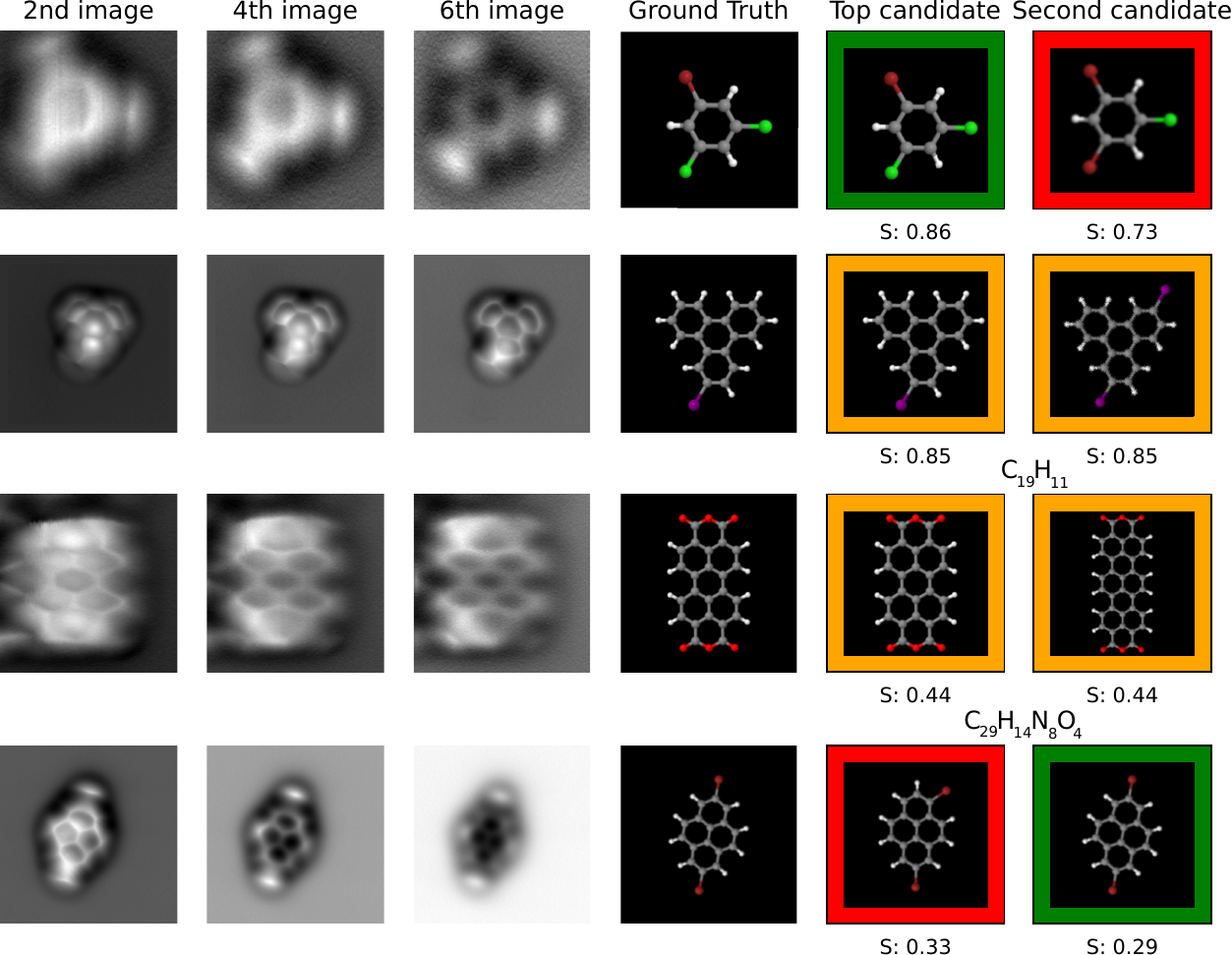}
    \caption{Chemical identification on experimental images. Molecules from first to last row are 1-Bromo-3,5-dichlorobenzene~\cite{Foster2022ACSNano} (CID: 29766), 2-iodotriphenylene~\cite{martin2019bond} (CID: 88955426), PTCDA~\cite{Foster2022ACSNano} (3,4,9,10-Perylenetetracarboxylic dianhydride
 CID: 67191) and 2,7-Dibromopyrene~\cite{Zhong2021Nov} (CID: 13615479).  The predicted chemical formula correctly solves the tie for the 2-iodotriphenylene molecule, predicting the GT molecule, but fails for PTCDA.  In all cases, we extract meaningful chemical information from the experimental image stack.\label{fig:exp_img_identification}.}  
\end{figure*}

In previous sections, we demonstrated that our strategy for molecular identification, combining the fingerprint and chemical formula models, works exceptionally well for simulated images (as illustrated in Fig.~\ref{fig:top_1acc_chemical_formula_corrugation}), achieving, on our large test set, a retrieval accuracy of 97.59\%. 
In Fig.~\ref{fig:exp_img_identification}, we benchmark our model over a limited set of  experimental cases: 1-Bromo-3,5-dichlorobenzene (CID: 29766)~\cite{Foster2022ACSNano},  2-iodotriphenylene (ITP, CID: 88955426)~\cite{martin2019bond}, PTCDA~\cite{Foster2022ACSNano}   
and 2,7-Dibromopyrene~\cite{Zhong2021Nov} (CID:13615479).  2-Iodotriphenylene was adsorbed on a Ag(111) surface while the rest of the molecules were deposited on Cu(111). Figures S4 and S5 show the complete stack of 10 constant--height images measured in the experiments. 
These experimental images clearly display the changes in the molecular configuration induced by the interaction with the substrate that we have already discussed from a theoretical perspective in section.~\ref{sec:corrugation}.  For example, HR--AFM images for PTCDA on Cu(111) in the third row of Figure~\ref{fig:exp_img_identification} clearly show a brighter contrast on the left side of the molecule, at variance with the symmetry that we could expect from the gas--phase structure. This effect stems from the non--planar adsorption of the molecule to the substrate.

 A key point when applying our model to experimental images and assessing its accuracy is the height range on which the molecules are imaged. In our dataset, the tip--sample distance ranges from 280 to 370 pm. This range, where the interaction changes from being slightly attractive to strongly repulsive, covers the typical imaging conditions. In experiments,  the height range explored is determined with respect to a specific set point (the position of maximum approach or where a referenced value of the tunneling current is measured by STM), but the absolute tip--sample distance is not known. 
Figures S4 and S5 compare the experimental image stacks used in Fig.~\ref{fig:exp_img_identification} to their corresponding simulations, with the same method our dataset~\cite{QUAM-AFM_JCIM} was generated. From the comparison, we can conclude that, for 1-Bromo-3,5-dichlorobenzenethe, experiments are exploring in a tip--sample distance range similar to the one considered in the training of the model, while, for ITP (with an experimental range of 72~pm) and PTCDA, images are sampled much closer than what the model expects based on the training data. For 2,7-Dibromopyrene, the experimental range is 135pm, 45 pm greater than the theoretical range of 90 pm. Our model has generalized to distances outside its training data to correctly predict the fingerprints of the molecule. 

Despite the differences in the tip height range sampled in some of the experiments and the internal corrugation induced by the substrate, the model is able to generalize and provide meaningful information about the chemical composition and bonding topology of the molecules. For 1-Bromo-3,5-dichlorobenzene (Fig.~\ref{fig:exp_img_identification}, 1st row), the  fingerprint model correctly identifies the molecule with a very high Tanimoto similarity, $S=0.86$. Notice that the model is capable of discriminating among the different halogen species, identifying the presence of two Cl and one Br atom, and retrieving the correct molecule.
For the ITP and PTCDA molecules, the fingerprint model arrives at a tie because the first and second candidates both have the same fingerprints. In the case of PTCDA, it is rather remarkable that the model is able to retrieve the fingerprints (although with a low Tanimoto similarity S=0.44) from the low quality of the experimental image.
In both cases, the tie stands from the fact that the frequency (number of occurrences) of a certain substructure is removed from the fingerprints. 

In the case of ITP,  although the predicted chemical formula is not completely correct (predicted C$_{19}$H$_{11}$ vs the true chemical formula  C$_{18}$H$_{11}$I), it  provides enough information to break the tie and achieve molecular identification. This is not the case for PTCDA, where the prediction is  C$_{29}$H$_{14}$N$_{8}$O$_{4}$ while the true chemical formula is C$_{24}$H$_{8}$O$_{6}$.
In the last row, the low values for the Tanimoto similarity indicate that the model has problems retrieving the fingerprints. It correctly predicts the overall topology of the molecule and the presence of two bromine atoms, but interchanges the position of one of the Br atoms with a neighboring H atom. The chemical formula model correctly predicts the presence of the two Br atoms (predicted C$_{14}$H$_{7}$Br$_{2}$N vs the true chemical formula  C$_{16}$H$_{8}$Br$_{2}$), but in this case it is not useful to choose between the two top candidates.

In the case of ITP and 1-Bromo-3,5-dichlorobenzene, we found that small variations in the scan size and pixel resolution of the experimental images caused huge changes in the ability of the model to retrieve the molecular fingerprints, as shown by the changes in the Tanimoto (of the order of 0.4). This sensitivity was absent from both the remaining experimental images and the simulated images. 

We have tried to understand this sensitivity looking at the attention maps of the images generated using Grad-CAM~\cite{selvaraju2017grad} (see section~S8). After a careful exploration,  we found that, for the scan size and pixel resolutions where the model performs the best, the model is paying more attention to the regions where the heteroatoms are located (Fig.~S6). Although further work, exploring systematically more experimental cases,  is clearly needed, these two examples suggest that attention maps, that do not require  any other input as the ground truth fingerprints,  should provide a powerful protocol for the validation of the model's predictions on experimental images.

To conclude, the fingerprint  model shows a very promising performance, while the results from the chemical formula model are more modest but good enough in some cases to break the ties associated with the loss of information in the construction of the fingerprints. Despite these good results, a larger, systematic analysis with proper experimental data is necessary to further address the accuracy of our model.

\section{Conclusions}\label{sec13}

A pipeline for automated molecular identification has been presented in this study. The pipeline predicts both molecular structure and chemical composition from HR-AFM image stacks. To achieve this, a convolutional neural network was trained using the QUAM--AFM dataset. The network retrieves the molecular fingerprint, ECFP4, with high accuracy, 0.95 median Tanimoto similarity in the test set. This accuracy is attributed to the choice of molecular descriptor. ECFP4 captures a lot of structural information and, unlike other codifications such as IUPAC names, SMILES, or SELFIES, ECFP4 uses binary vectors, making their prediction the well-studied problem of multilabel classification. Knowledge of a molecule's fingerprints has a wide range of applications. Designed for high-throughput screening, these fingerprints are particularly good at encoding the presence or absence of specific substructures. Beyond molecular identification, they can be useful for other downstream tasks, such as predicting quantum mechanical properties~\cite{Wu2018Jan}, thermodynamic properties~\cite{Besel2023Jul} and even finding new antibiotics with specific properties~\cite{Wong2024Feb}.  
We have shown how it is possible to determine the molecule among a list of candidates by a virtual screening process done by ranking the possible candidates by decreasing order of Tanimoto similarity. 
To compensate the loss of  the frequency of the identified substructures during the hashing process, we can re--rank the final candidates using another CNN designed to predict the chemical formula, boosting the accuracy of the prediction up to a 97.6\%

Although trained and tested with simulated HR--AFM images, the final goal of our model is to retrieve the molecular fingerprints and achieve molecular identification from experimental images. To that end, we have proved that our model can distinguish chemical contrast from the structural changes induced by molecular adsorption and performed few identification  tests with experimental images that have shown very promising results.

A systematic collaboration between theory and experiment is needed to further develop the model to work under real experimental conditions. Particularly promising in this direction is the possibility to use attention maps to improve and to validate of the models' predictions on experimental images. Even with  its current limitations, 
our model provides an accurate, straightforward method for automated molecular identification that can boost the chemical analysis and characterization of complex molecular materials such as intermediates and products of on-surface reactions, soot molecules, fuel pyrolysis products, dissolved organic carbon, or other petroleum products as well as materials of interest for catalysis or astrochemistry.

\section*{Acknowledgments}

We thank the Adam Foster's and Peter Liljeroth's groups for making the experimental images of the 1-Bromo-
3,5-dichlorobenzene and PTCDA molecules publicly available. 
We acknowledge support from the Spanish Ministry of Science and Innovation, through project PID2020-115864RB-I00, TED2021-132219A-I00,  and the ``Mar\'{\i}a de Maeztu'' Programme for Units of Excellence in R\&D (CEX2023-001316-M). We thank Sebastian Ahles and Hermann A. Wegner for providing the 2-iodotriphenylene molecules and Daniel Martin-Jiminez for performing the LT-AFM measurements. We acknowledge partial funding by the Deutsche Forschungsgemeinschaft (DFG) via grants EB 535/1-1, EB 535/4-1, SCHI 619/13 and the LOEWE Program of Excellence of the Federal State of Hesse via the LOEWE Focus Group PriOSS ``Principles of On-Surface Synthesis''.

\bibliographystyle{naturemag}

\bibliography{bibliography}

\end{document}


\maketitle

\section{Initialization of the Fingerprint and Chemical formula models}
\label{subsection:models_initialization}
Here, we provide information about the choice for the initial parameters in the two models.
Since in our adaptation of EfficientNet-B0~\cite{efficientnet}, we change the first convolution from 3 color channels to 10, we are increasing the number of parameters in this layer. Since we know from  Yosinski et al.~\cite{Yosinski2014Nov} and Zeiler and Fergus~\cite{Zeiler2013Nov} that filters in the initial layers of Convolutional Neural Networks (CNNs) often correspond to low-level representations such as edges and color blobs, we have initialized our model effectively by replicating the weights from the original Conv2D layer across the new channels in the modified EfficientNet-B0 architecture. \\

\begin{figure}[b!]
    \centering
    \includegraphics[width = 1\linewidth]{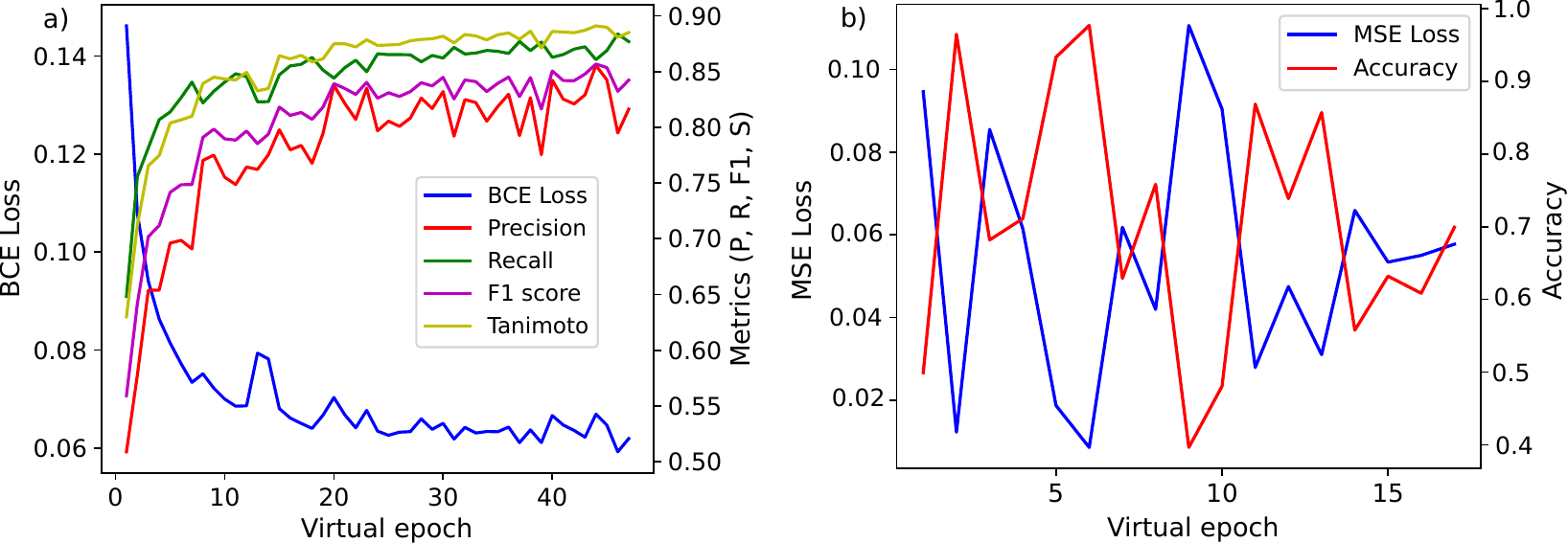}
    \caption{Validation losses and metrics for both {\bf a)} the molecular fingerprints and {\bf b)} the chemical formula models. In {\bf a)}, the monitored metrics are the precision, recall, F1 score and Tanimoto similarity, the last one being the most important for chemical identification. There are diminishing returns in training after epoch 10. For {\bf b)},  since we start from the pre-trained backbone from the molecular fingerprints model, the model almost immediately reaches a very good performance (MSE loss $<0.1$ atoms). The loss oscillates during training, indicating that we reached a stable local minima.}
    \label{fig:joint_training_metrics}
\end{figure}

Another trick for accelerating convergence of the model is to initialize the last layer's bias weights with a prior. For the fingerprints model, this prior is the logit of the probability of appearance of each substructure in the training set. For the chemical formula model, we have chosen the mean count of each atom in the training set.

\section{Training hyperparameters}
\label{subsection:training_hyperparameters}

\begin{table*}[t!]
\centering
\begin{tabular}{|l|c|c|}
\hline
\textbf{Training Parameter} & \textbf{Molecular Fingerprint Model} & \textbf{Chemical Formula Model} \\
\hline
Optimizer                   & Adam                                & Adam                            \\
\hline
Learning rate               & $1 \times 10^{-3}$                  & $\sqrt{2} \times 10^{-3}$       \\
\hline
Batch size                  & 50                                  & 100                              \\
\hline
Loss function               & Binary Cross Entropy                                 & Mean Squared Error                              \\
\hline
\end{tabular}
\caption{Training parameters for the molecular fingerprint and the chemical formula models. \label{table:training_params}}
\end{table*}

\begin{table*}[h!]
\centering
\begin{tabular}{|l|c|c|}
\hline
\textbf{Augmentation Method} & \textbf{Probability} & \textbf{Augmentation Range (Unit)} \\
\hline
Rotation                     & 0.5                   & [-180, 180) (degrees)            \\
\hline
Zoom (in/out)                & 0.7                   & [0.3, 1.7]                         \\
\hline
Translation (vertical/horizontal) & 0.3              & [-20, 20] (pixels)                \\
\hline
Shear                         & 0.3                  & [-10, 10] (degrees)              \\
\hline
Gaussian Noise                & 1                    & $\mu=0$, $\sigma=2$            \\
\hline

\end{tabular}
\caption{Data augmentation parameters for the training of both models. For each HR-stack, the same rotation, zoom translation and shear transformations are applied to all the constant--height images, while the gaussian noise is applied individually to each image. \label{table:data_aug_params}}
\end{table*}

Both the molecular fingerprint and chemical formula models employ the Adam optimizer with the parameters shown in Table~\ref{table:training_params}.  Learning rates were set to $1 \times 10^{-3}$ for the molecular fingerprint model and $\sqrt{2} \times 10^{-3}$ for the chemical formula model as we doubled the batch size for the chemical formula model (Table~\ref{table:training_params}). Finally, we select the loss functions to be Binary Cross Entropy for the molecular fingerprint model and Mean Squared Error for the chemical formula model as they are the most commonly used for multilabel classification and regression respectively. Figure~\ref{fig:joint_training_metrics} displays the validation losses and metrics for both models, showing that 10 epochs are enough for the training of  fingerprint model, while, for the chemical formula,  since we start from the pre-trained backbone from the molecular fingerprints model, the model almost immediately reaches a very good performance (MSE loss $<0.1$ atoms). 

\begin{figure*}[ht]
    \centering
    \includegraphics[width = 0.9\linewidth]{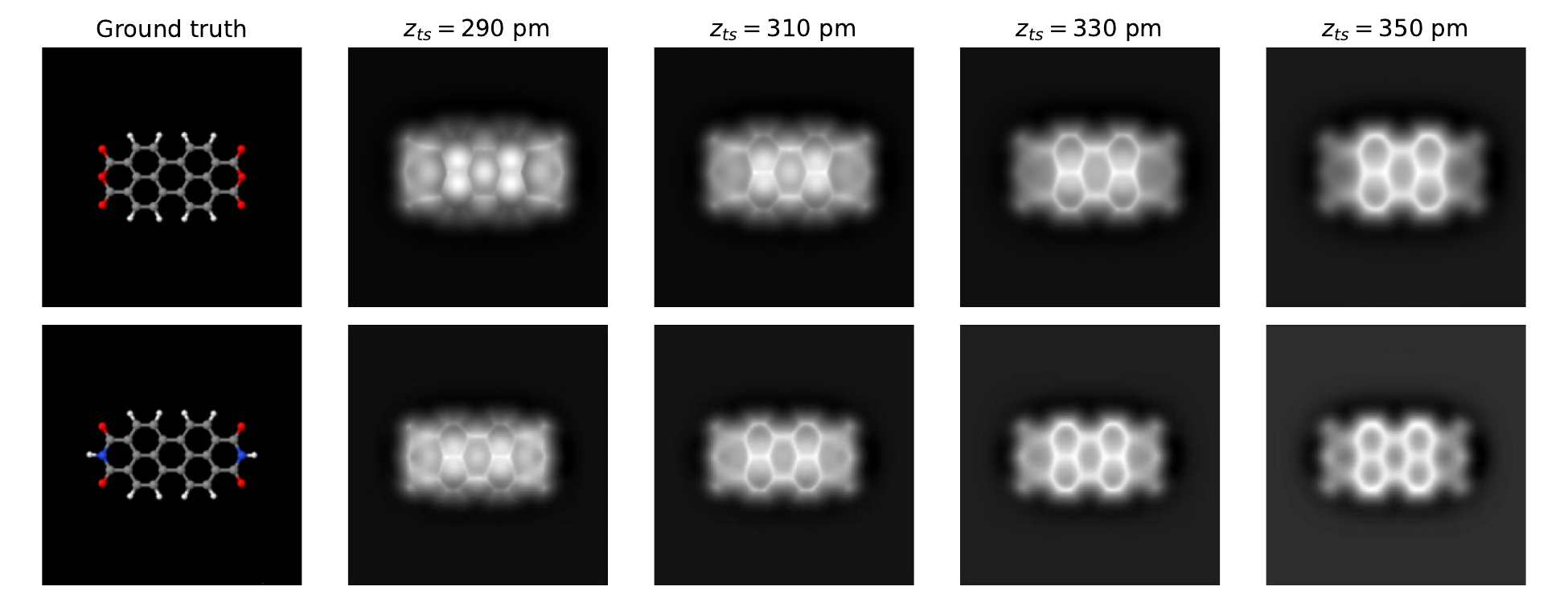}
    \caption{Simulated image stacks of PTCDA adsorbed on Ag(111) (first row) and Perylimid (CID: 66475) gas phase. The visual features of the AFM images corresponding to the oxygens of PTCDA and to the NH groups are very similar.
     \label{fig:ptcda_ag_and_nh}}  

\end{figure*}

\section{Balancing precision and recall during training}
\label{section:p_c_parameter}

The $p_c$ parameter in the Binary Cross Entropy loss function introduces a trade--off between precision and recall, where $p_c<1$ increases precision and $p_c>1$ increases recall.
Usually, we choose this weight to be the negative cases divided by the positives and carry out the computation for each label (each bit on the molecular fingerprint).

We observed that the values of $p_c$ were very high using this method, resulting in recalls too big, what decreased the F1 score. To improve the precision in these labels, we applied a penalty to values of $p_c$ greater than 1:
\begin{equation}
p_c' = 
\begin{cases} 
1 + \frac{p_c-1}{10} & \text{if } p_c > 1 \\
p_c & \text{if } p_c \leq 1 
\end{cases}
\end{equation}
This penalty marginally improved the F1 metric, while it had little effect on the Tanimoto similarity, which is the most important metric in this study.

\section{Data augmentations}
\label{subsection:data_augmentations}
We incorporate several data augmentation with two intentions:
\begin{itemize}
    \item Increase the number of different training samples.
    \item Make the model robust against experimental conditions.
\end{itemize}
There are several sources of noise in an experiment: brownian motion of the tip due to thermal noise, asymmetrical tips, drift, surface tilt of the molecule. etc.
Since the QUAM-AFM images are simulated, they do not reflect any of these conditions, so we need to add them in the form of data augmentations: rotations, translations, zooming, shear and gaussian noise augmentations are performed during training. The probability of occurrence and intensity of each augmentation is provided in Table~\ref{table:data_aug_params}.
\begin{figure*}[hb]
    \centering
    \includegraphics[width = 0.8\linewidth]{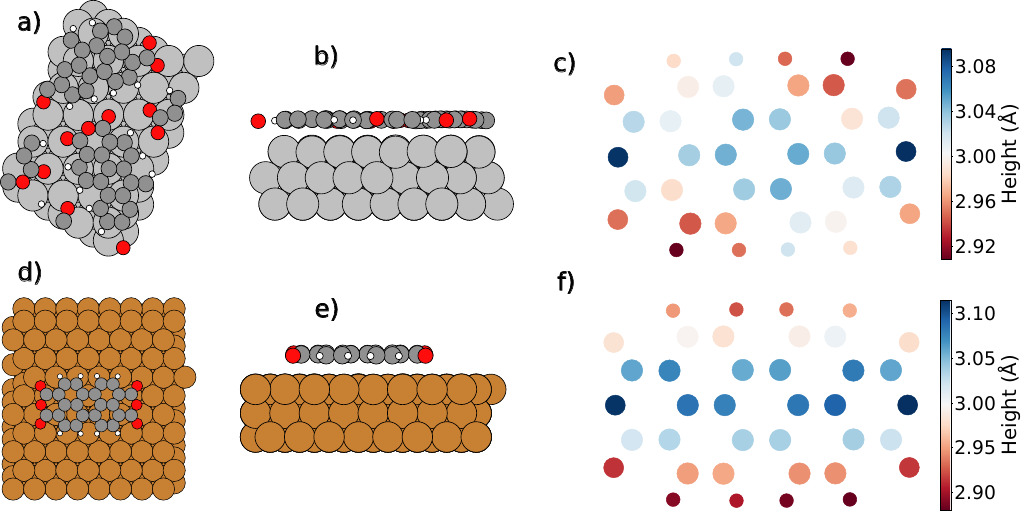}
    \caption{Visualization of the PTCDA molecule adsorbed on Ag(111) and Cu(111) surfaces. Panels a) and d) show the top view, while b) and e) present the front view of the PTCDA molecule adsorbed on Ag111 and Cu111, respectively. Panels c) and f) depict the height distribution of the relaxed PTCDA molecule on Ag111 and Cu111, illustrating the surface effect on the molecular corrugation.
     \label{fig:ptcda_ag_cu_sketch}}  
\end{figure*}

\section{Atom count model}
\label{subsection:atom_count_model}

To overcome the limitation posed by the hashing step in the generation of the chemical fingerprints, we have developed a second model for atom count regression that we use to reorder top candidates. This new CNN shares the same backbone of the molecular fingerprints network and only changes the last layer to be a regression head with 10 neurons, one for each atomic species. Its performance is outstanding, correctly predicting the atom frequency in most of the samples from the test set (Table~\ref{table:combined_classification_regression_metrics}), the only exception been the case of P atoms. We attribute this to the limited number of molecules containing this atom in the training set.

\begin{table*}[h!]
\centering
\begin{tabular}{ccccccc}
\multicolumn{4}{c}{Classification Metrics} & \multicolumn{3}{c}{Regression Metrics} \\
\hline
Atom & Precision & Recall & F1 Score & MAE & Pearson's r \\
\hline
C & 1.0000 & 1.0000 & 1.0000 & 0.0115 & 0.9996 \\
Br & 0.9994 & 0.9979 & 0.9986 & 0.0007 & 0.9983 \\
Cl & 0.9986 & 0.9991 & 0.9989 & 0.0008 & 0.9988 \\
F & 0.9985 & 0.9975 & 0.9980 & 0.0015 & 0.9971 \\
I & 0.9959 & 0.9973 & 0.9966 & 0.0004 & 0.9954 \\
N & 0.9994 & 0.9993 & 0.9994 & 0.0089 & 0.9977 \\
O & 0.9991 & 0.9991 & 0.9991 & 0.0058 & 0.9978 \\
P & 0.8884 & 0.6782 & 0.7692 & 0.0005 & 0.7854 \\
S & 0.9980 & 0.9973 & 0.9976 & 0.0018 & 0.9962 \\
H & 0.9999 & 1.0000 & 1.0000 & 0.0121 & 0.9993 \\
\end{tabular}
\caption{Atom Classification and Regression Metrics. Computed on the approximately 275k molecules in the test set. We achieve nearly perfect accuracy for all atoms but phosphorus, which gives worse results because it's underrepresented in the molecules from QUAM--AFM dataset.}
\label{table:combined_classification_regression_metrics}
\end{table*}

\begin{figure*}[!ht]
    \centering
    \includegraphics[width = 0.67\linewidth]{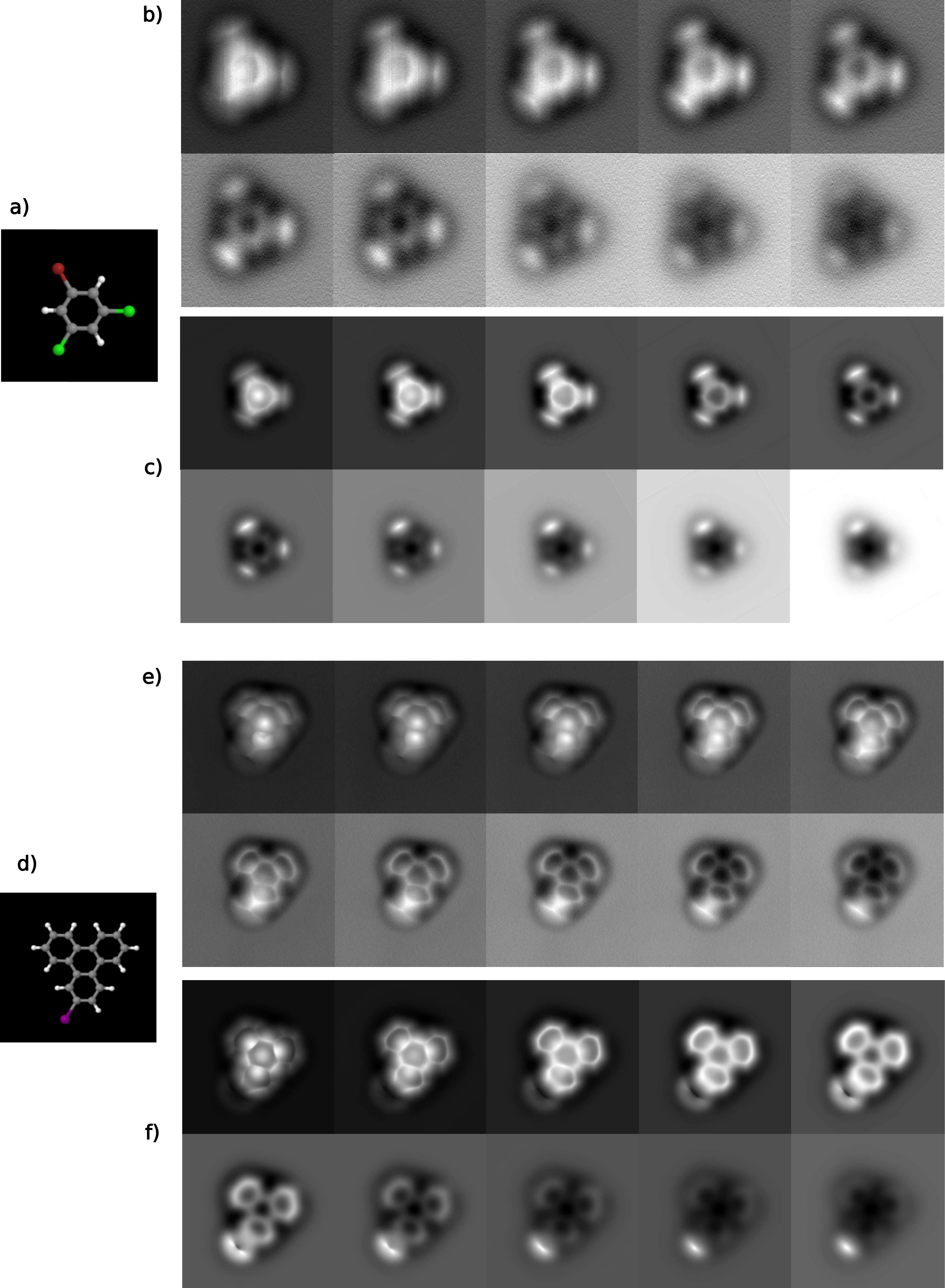}
    \caption{Comparison of experimental and images and corresponding theoretical simulations.  Jmol images (a,d) experimental (b,e) and simulated (c,f) image stacks for 1-Bromo3,5-dichlorobenzene (CID: 29766) and 2-iodotriphenylene (CID: 88955426) respectively. The operational parameters for the simulations were set to 140 pm for the amplitude and 0.4 N/m for the CO elastic constant in (c), and to 40 pm for the amplitude and 0.4 N/m for the CO elastic constant (f).}
    \label{fig:exp_no_zoom_theory_bcb_2iodotriphenylene}
\end{figure*}

\begin{figure*}[!ht]
    \centering
    \includegraphics[width = 0.67\linewidth]{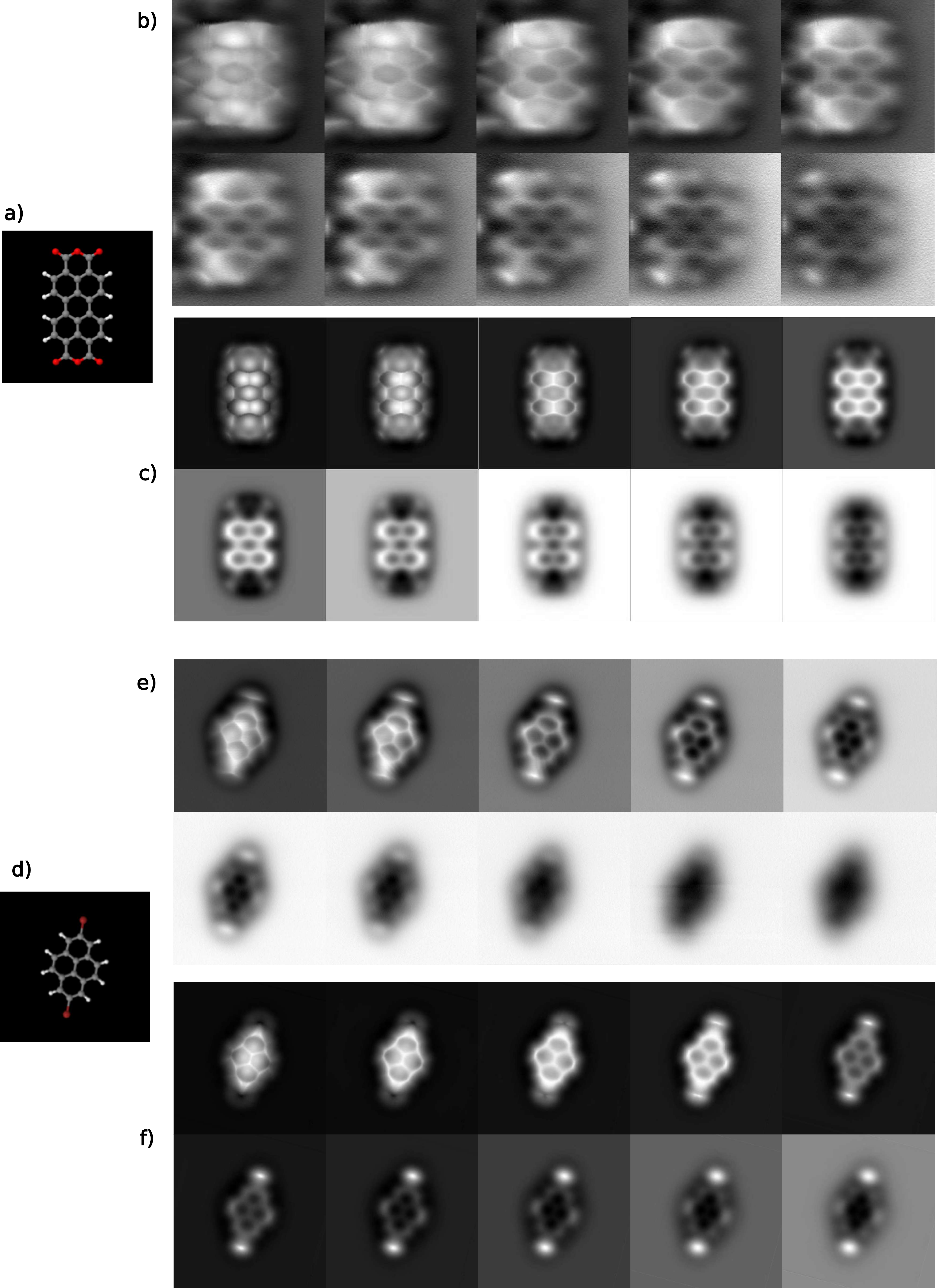}
    \caption{Comparison of experimental and images and corresponding theoretical simulations.  Jmol images (a,d) experimental (b,e) and simulated (c,f) image stacks for PTCDA (3,4,9,10-
Perylenetetracarboxylic dianhydride CID: 67191) and 2,7-Dibromopyrene (CID: 13615479) respectively. The operational parameters for the simulations were set to 40 pm for the amplitude and 0.4 N/m for the CO elastic constant in (c), and to 40 pm for the amplitude and 1 N/m for the CO elastic constant (f).}
    \label{fig:exp_no_zoom_theory_ptcda_dibromopyrene}
\end{figure*}

\section{Surface--induced corrugation}
\label{subsection:surface_induced_corrugation}
Fig.~\ref{fig:ptcda_ag_and_nh} compares some of the AFM images in the simulated 3D stacks for PTCDA adsorbed on silver (first row) and Perylimid (CID: 66475) gas phase. When the PTCDA molecule is adsorbed on Ag(111), the middle oxygen atom gets pushed away from the surface, which enhances its contrast on the images. This surface--induced corrugation makes the model choose the Perylimid molecule as top candidate with high confidence. 
Fig.~\ref{fig:ptcda_ag_cu_sketch} shows the adsorption geometry for PTCDA on Ag(111) and Cu(111) surfaces. The (c) and (d) panels highlight the surface induced corrugation on the PTCDA molecule, which is completely flat in gas phase.

\section{Comparing simulated and experimental images}

Figures~\ref{fig:exp_no_zoom_theory_bcb_2iodotriphenylene} and~\ref{fig:exp_no_zoom_theory_ptcda_dibromopyrene} present the experimental image stacks used in this article to test our model, along with the corresponding simulations. The operational parameters for the simulations (oscillation amplitude and lateral stiffness of the CO molecule) have been chosen, among those available in the QUAM--AFM dataset, to better reproduce the experimental contrast. Notice that, in almost all cases, the height and the relative scale of the molecule compared to the frame in the experimental images differ from what would be obtained using the method of simulation of the dataset~\cite{QUAM-AFM_JCIM}.

\begin{figure*}[ht]
    \centering
    \includegraphics[width = 0.7\linewidth]{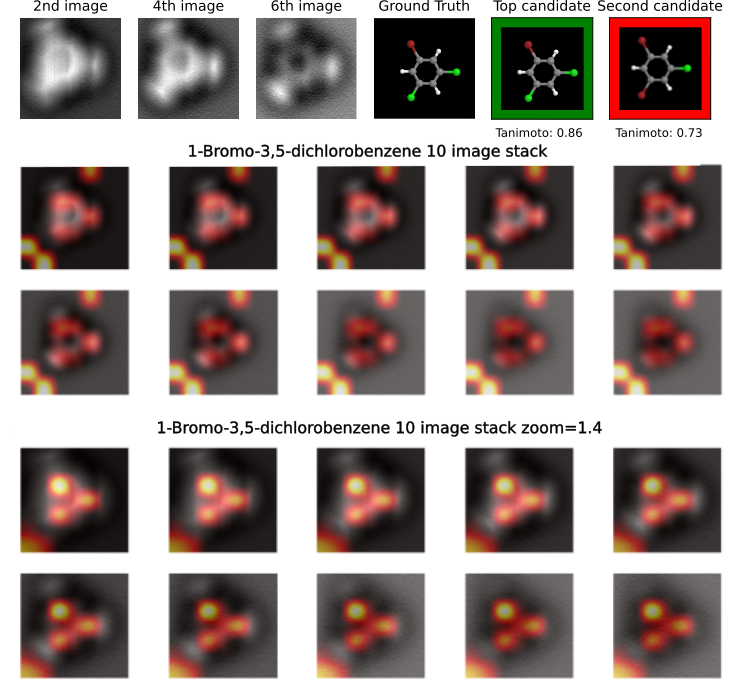}
    \caption{Grad-CAM algorithm applied to the BCB molecule for different zooms. The model pays more attention in the images with the larger zoom to the heteroatoms, which are key for correctly predicting the correct chemical fingerprints.}
    \label{fig:gradcam_figure_experimental_bcb}
\end{figure*}

\section{Attention maps}

\label{subsection:explainability}
In the identification of 2-iodotriphenylene and 1-Bromo-3,5-dichlorobenzene, we found that the Tanimoto similarity retrieved by our model varied significantly when small changes in the pixel resolution, and therefore in the scan size, were applied to the experimental images. As this sensitivity was not observed for theoretical images, we decided to check the attention maps produced by the model at several pixel resolutions using the Grad-CAM algorithm~\cite{selvaraju2017grad}. 

To adapt Grad-CAM to a multilabel classification, we compute the attention map of the molecule for each predicted "on" label $c$ using the standard Grad-CAM implementation. Each resulting heat-map $L^c$ is normalized individually:
\begin{equation}
L^{c}_{norm} = \frac{L^{c}_{i,j}}{max(L^{c}_{i,j})}
\end{equation}
Finally, the mean of all normalized heat-maps yields the final heat-map per molecule:
\begin{equation}
L_{final} = \frac{1}{C} \sum_{c=1}^{C} L^{c}_{norm}
\end{equation}
where $C$ represents the total count of predicted "on" labels. 
With Grad-CAM, we generate a 2D-heat-map of the AFM image stack, where the intensity on each region corresponds to its importance in predicting the molecular fingerprint.\\
Fig.~\ref{fig:gradcam_figure_experimental_bcb} illustrates, for the case of 1-Bromo-3,5-dichlorobenzene (BCB), how decreasing the scan size by a factor of 1.4 (zooms greater than one means dividing the scan size by the same factor) yields stronger attention over the heteroatoms and the structure in general, explaining the significant change in the Tanimoto similarity obtained with our model.

\bibliographystyle{naturemag}
\bibliography{bibliography}